\documentclass[sigconf, screen, arxiv]{acmart}

\usepackage{array}
\usepackage{makecell}
\usepackage{multirow}
\usepackage{pifont}
\usepackage{subcaption}
\begin{document}

\title{ALFEE: Adaptive Large Foundation Model for EEG Representation}

\author{Wei Xiong, \quad Junming Lin, \quad  Jiangtong Li, \quad  Jie Li, \quad  Changjun Jiang}
\affiliation{
  \institution{Department of Computer Science and Technology, Tongji University}
  \state{Shanghai}
  \country{China}}
\email{{xw1216, jiangtongli}@tongji.edu.cn}

\renewcommand{\shortauthors}{Xiong et al.}

\begin{abstract}
While foundation models excel in text, image, and video domains, the critical biological signals, particularly electroencephalography~(EEG), remain underexplored.
EEG benefits neurological research with its high temporal resolution, operational practicality, and safety profile.
However, low signal-to-noise ratio, inter-subject variability, and cross-paradigm differences hinder the generalization of current models.
Existing methods often employ simplified strategies, such as a single loss function or a channel-temporal joint representation module, and suffer from a domain gap between pretraining and evaluation tasks that compromises efficiency and adaptability.
To address these limitations, we propose the \textbf{A}daptive \textbf{L}arge \textbf{F}oundation model for \textbf{EE}G signal representation (\textbf{ALFEE}) framework, a novel hybrid transformer architecture with two learning stages for robust EEG representation learning.
ALFEE employs a hybrid attention that separates channel-wise feature aggregation from temporal dynamics modeling, enabling robust EEG representation with variable channel configurations.
A channel encoder adaptively compresses variable channel information, a temporal encoder captures task-guided evolution, and a hybrid decoder reconstructs signals in both temporal and frequency domains.
During pretraining, ALFEE optimizes task prediction, channel and temporal mask reconstruction, and temporal forecasting to enhance multi-scale and multi-channel representation.
During fine-tuning, a full-model adaptation with a task-specific token dictionary and a cross-attention layer boosts performance across multiple tasks.
After 25,000 hours of pretraining, extensive experimental results on six downstream EEG tasks demonstrate the superior performance of ALFEE over existing models.
Our ALFEE framework establishes a scalable foundation for biological signal analysis with implementation available at \url{https://github.com/xw1216/ALFEE}.

\end{abstract}

\ccsdesc[500]{Computing methodologies~Artificial intelligence}
\ccsdesc[500]{Computing methodologies~Cognitive science}
\ccsdesc[300]{Human-centered computing~HCI design and evaluation methods}

\keywords{Foundation Model, Multi-task Learning, EEG, Pretraining}

\maketitle

\section{Introduction}
The emergence of foundation models and their evolution into multimodal systems~\cite{devlin2019bert,achiam2023gpt,girdhar2023imagebind,betker2023improving,pagnoni2024byte,rasul2023lag} has transformed the artificial intelligence to process and integrate diverse information modalities. 
Specifically, multimodal foundation models exhibit strong capabilities in knowledge integration and data analysis while narrowing the gap between computational systems and real-world understanding, thereby driving advances in human productivity. 
Pioneering frameworks such as CLIP~\cite{radford2021learning} and MAE~\cite{he2022masked} have set new benchmarks in visual-semantic understanding, while BEATs~\cite{chen2023beat} has emerged as a paradigm-shifting approach for music audio feature extraction. 
Despite these achievements, current foundation models primarily focus on conventional modalities (\emph{e.g.}, text, image, video)~\cite{dosovitskiy2020image,alayrac2022flamingo,rombach2022high,wu2024nextgpt}, with their architectural derivatives largely confined to these domains. 
This leaves critical biological signal modalities, \emph{i.e.}, electroencephalography~(EEG), significantly underexplored in the era of foundation models. 
In particular, multi-channel time-series data like EEG offers vital insights into brain function and structure, which is essential in cognitive science.

\begin{figure}[t]
  \centering
  \includegraphics[width=0.93\linewidth]{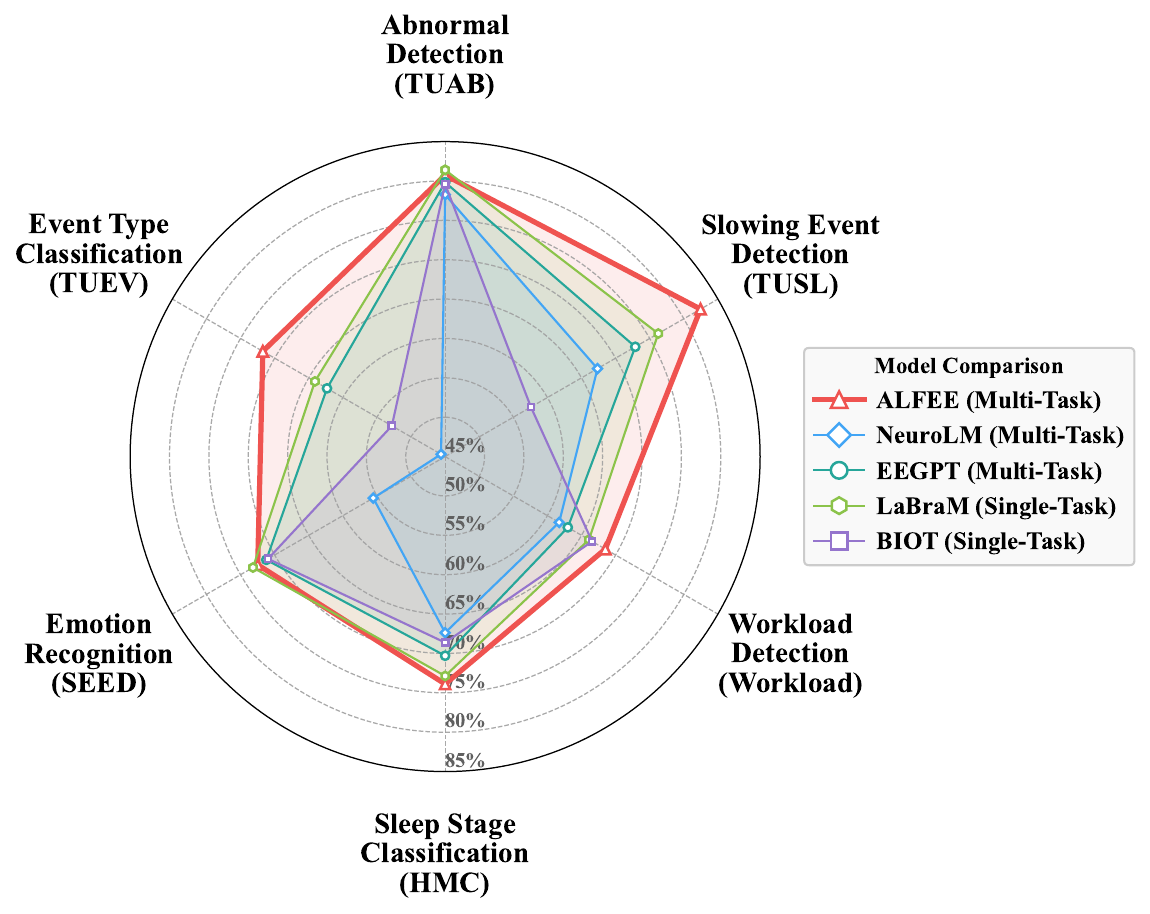}
  \vspace{-10pt}
  \caption{Balanced accuracy performance on six datasets.}
  \label{fig:radar}
  \vspace{-20pt}
\end{figure}

EEG, as a widely adopted non-invasive neural recording modality, offers unique advantages for both neurological research and medical applications~\cite{song2024eeg,gong2023astdf,li2022multi,demir2021eeg,xu2024wsel}.
Specifically, EEG exhibits \textbf{high temporal resolution} (superior to Near-Infrared Spectroscopy), \textbf{operational practicality} (more cost-effective than functional Magnetic Resonance Imaging), and \textbf{safety profile} (non-invasive compared to Positron Emission Tomography).
These advantages render EEG indispensable for detecting neurological anomalies, such as epileptic seizures, sleep disorders, and traumatic brain injuries~\cite{kumari2024brain}.

Recent advances in EEG signal modeling have introduced several innovative pretraining frameworks with distinct methodological contributions. LaBraM~\cite{jiang2024large} utilizes vector-quantized neural spectrum prediction to establish a comprehensive semantic tokenization system during the pretraining phase, while its neural transformer architecture facilitates simultaneous learning of spatiotemporal characteristics inherent in EEG signals.
The EEGPT framework~\cite{wang2024eegpt} introduces a dual self-supervised paradigm combining spatiotemporal representation alignment with mask-based reconstruction. This hierarchical architecture initially derives stable spatial features from short-term EEG segments, subsequently modeling temporal dependencies across extended EEG signal sequences.
Notably, NeuroLM~\cite{jiang2024neurolm} presents an integrated approach by combining the LaBraM encoder with the GPT-2 architecture~\cite{radford2019language}, effectively harnessing the representational capacity of large-scale language models. This hybrid architecture demonstrates enhanced performance in multimodal learning scenarios, particularly in tasks requiring simultaneous processing of multiple cognitive objectives.

Despite significant advancements in EEG signal representation, EEG foundation models continue to encounter several critical challenges.
First, variations in channel counts across different datasets pose challenges in efficiently learning channel representations, as they must adhere to varying standards within a unified framework.
Second, EEG signals are characterized by two primary dimensions: the channel dimension, which reflects the activity of various brain regions at a specific moment, and the temporal dimension, which captures the evolution of these signals over time.
Current methods typically rely on either a single loss function or a joint representation module, which limits their capability to fully capture the complex, dimension-specific features.
Third, a significant domain gap persists between upstream pretraining tasks and downstream evaluation tasks.
Existing methods either employ single-task fine-tuning for specific applications or freeze pretrained parameters while incorporating additional modules for downstream tasks, which reduces adaptability and impairs generalization and robustness.

In this work, we present the \textbf{A}daptive \textbf{L}arge \textbf{F}oundation model for \textbf{EE}G signal representation (\textbf{ALFEE}) framework, which is built on a hybrid attention architecture and employs two optimization stages to address the above issues.
From the framework perspective, our ALFEE separates channel-wise feature aggregation from temporal dynamics modeling by first employing a channel encoder that adaptively compresses variable channel information to accommodate diverse channel counts.
Subsequently, a temporal encoder captures task-guided temporal evolution.
Finally, a decoder using hybrid attention with a learnable recovery query reconstructs the EEG signal in both the temporal domain and the frequency domain.
From the training perspective, our framework optimizes four tasks during the pretraining stage: task prediction, channel mask reconstruction, temporal mask reconstruction, and temporal forecasting, effectively enhancing its multi-channel and multi-scale representation capabilities.
In the fine-tuning stage, a full-model fine-tuning strategy is employed, utilizing a task-specific token dictionary augmented by a cross-attention layer, and comprehensively improving performance in multi-task scenarios.

To validate the effectiveness of our framework, we pretrain ALFEE with four model sizes on a large-scale EEG dataset, comprising over 25,000 hours and 15 datasets spanning 8 tasks~\cite{jiang2024neurolm}.
Next, we finetune ALFEE with different model sizes on 6 downstream tasks, including \textbf{TUAB}~\cite{7405423}, \textbf{TUEV}~\cite{7405421}, \textbf{TUSL}~\cite{8257018}, \textbf{SEED}~\cite{7104132}, \textbf{HMC}~\cite{10.1371/journal.pone.0256111}, and \textbf{Workload}~\cite{data4010014}.
Extensive experimental results~(Figure~\ref{fig:radar}) demonstrate the superior performance of ALFEE over existing multi-task state-of-the-art~(SOTA) EEG foundation models~(\emph{i.e.}, NeurLM and EEGPT), showcasing its advanced capability in robust and generalized feature representations from noisy EEG signals.
Finally, further experiments confirm the contributions of each loss and model size within the framework, validating the scaling-law in EEG signal representation.
Our contributions are summarized as

\begin{itemize}
\item \textbf{Hybrid Attention Architecture}: An hybrid mechanism with self- and cross-attention that separates channel-wise feature aggregation from temporal dynamics modeling, enabling EEG representation with variable channel setting.
\item \textbf{Multi-Task, Multi-Channel, Multi-Scale Pretraining}: A unified training framework that combines: a) temporal forecasting; b) channel masked reconstruction; c) temporal masked autoencoding; and d) task prediction, enhanced by Power Spectral Density (PSD) features for frequency domain.
\item \textbf{Adaptive Representation Learning}: A modular architecture comprising a channel encoder, a temporal encoder, and an EEG decoder that adapts to heterogeneous EEG signals and enhances the capability in robust representations.
\item \textbf{Extensive Experiments}: Extensive experimental results on a pretraining dataset comprising over 25,000 hours EEG signals and 6 evaluation datasets, demonstrating the superior performance of ALFEE over existing models and further validating the scaling law in EEG signal representation.
\end{itemize}

\section{Methodology}
\begin{figure*}[t]
  \centering
  \includegraphics[width=\textwidth]{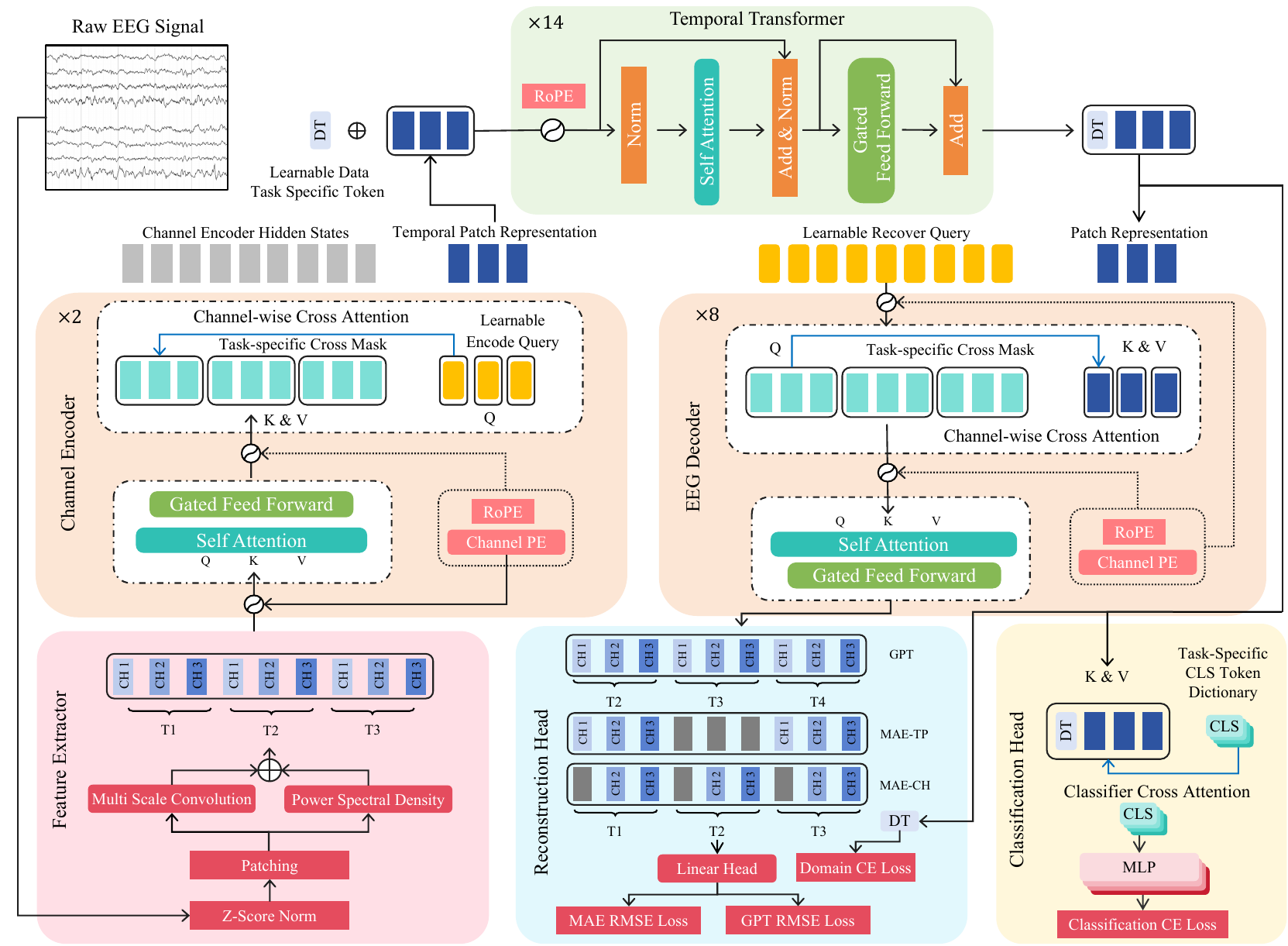}
  \caption{The overall architecture of ALFEE. (1) Feature Extractor: processes raw EEG signals via multi-scale convolution and PSD analysis; (2) Channel Encoder: captures channel-wise dependencies using cross-attention mechanisms; (3) Temporal Encoder: processes temporal feature through stacked transformer layers, guided by learnable task token; (4) EEG Decoder: reconstructs signals through attention-based refinement; (5) Pretraining Head: learns signal reconstruction via complicated loss minimization; and (6) Finetuning Head: gather information by task-specific token for downstream tasks. All the self- and cross-attention layers exploits the multi-head attention and different attention masking strategies are applied in each module.}
  \label{fig:model-architecture}
  \vspace{-10pt}
\end{figure*}

In this section, we present a detailed description of the architecture and optimization of the ALFEE model.
The model follows an encoder-decoder paradigm during pretraining, where an EEG feature encoder, composed of a composite feature extractor, a channel encoder, and a temporal transformer, first maps the input to a compact representation, and a decoder then reconstructs the input from this compact representation.
In the finetuning stage, a finetuning head is attached to temporal encoder for downstream classification.

\subsection{Preliminaries}\label{sec:prelimi}
To handle the issue that channel counts vary for different tasks and datasets, we first collect all existing electrode positions among different standards and predefine an electrode set $\mathbb{C} = {c_1, c_2, \ldots, c_{|\mathbb{C}|}}$, where $|\mathbb{C}| = 90$, containing all standard positions in the international 10-10 system in addition to T1, T2, A1, and A2~\cite{oostenveld2011fieldtrip}. 
For all involved datasets, the signals are resampled to $f_s=256 \, \text{Hz}$. 
The learnable query is denoted as $\mathbf{l_{*}}$.
The resampled EEG signal data is formulated as $\mathbf{x_0}\in \mathbb{R}^{C_0 \times P_0}$, where $C_0$ denotes the number of original electrodes in the dataset and $P_0$ denotes the number of sample points.
After slicing, the sample points are split into $n=C_0 \times \lfloor {P_0 / f_s} \rfloor$ non-overlapping one-second patches with patch size $P=f_s$.
Subsequently, EEG patch samples are denoted as $\mathbf{x_p} \in \mathbb{R}^{B \times T \times C \times f_s}$, where $B$, $T$, $C$, and $f_s$ denote the batch size, the number of time steps, the channel size, and the patch size after batching, respectively.

\subsection{Model Architecture}

ALFEE adopts a hierarchical architecture with the following components: 1) Feature Extractor; 2) Channel Encoder; 3) Temporal Encoder; 4) EEG Decoder; 5) Pretraining Head; and 6) Finetuning Head. 
As illustrated in Figure~\ref{fig:model-architecture}, ALFEE processes EEG signals with a time-frequency domain feature extractor, followed by encoder-decoder to capture EEG feature and reconstruct EEG signal.

\subsubsection{\textbf{Feature Extractor}} 

\quad\\\noindent\textbf{Input Normalization and Patching:}
To control the scale of the training loss and prevent gradient explosion in automatic mixed precision~\cite{micikevicius2017mixed}, we first apply z-score normalization to the resampled EEG signal data $\mathbf{x_0}$ along the temporal dimension:
\begin{equation}
    \hat{\mathbf{x}}\mathbf{_0} = {\mathbf{x_0} - \mu_\mathbf{x_0} \over \sigma_\mathbf{x_0} + \epsilon}
    \label{eq:znorm}
\end{equation}
where $\mu_{\mathbf{x_0}}$ and $\sigma_{\mathbf{x_0}}$ are calculated only on the temporal dimension, and $\epsilon = 1e^{-5}$ prevents division by zero.
Then we divide the input samples as described in Section~\ref{sec:prelimi} and organize them into mini-batches, leading to $\mathbf{x_p}\in\mathbb{R}^{B \times T \times C \times f_s}$ for further operation.

\noindent\textbf{Multi-Domain Feature Extraction:}
For the purpose of extracting informative EEG features in both temporal and frequency domains, we first implement multi-scale convolution using stacked \texttt{Conv1D} layers~\cite{masci2011stacked,szegedy2017inception} and compute the PSD using the Fast Fourier Transform~(FFT).
Temporal domain features describe transient characteristics, while frequency domain features illustrate global characteristics.
Given a normalized sample $\mathbf{x_p} \in \mathbb{R}^{B \times T \times C \times f_s}$, the tensor is first reshaped into $\mathbf{x^\prime_p} \in \mathbb{R}^{(B \cdot T \cdot C) \times 1 \times f_s}$.
The operation of the $l$-th \texttt{Conv1D} layer ($l \in {1, 2, \cdots, L}$) in the \texttt{StackConv} is defined as:
\begin{equation}
    \mathbf{H}^{(l)} = \mathrm{LayerNorm} \left( \mathrm{GELU}\left( \mathrm{Conv1D}\left(\mathbf{H}^{(l-1)} \right) \right) \right)
    \label{eq:stack-conv}
\end{equation}
where $\mathbf{H}^{(l)}$ denotes the representation at the $l$-th layer with $\mathbf{H}^{0}=\mathbf{x^\prime_p}$.
Then, the output ($\mathbf{H}^{(L)}$) is compressed along the patch dimension to a fixed length, followed by reshaping to restore the channel structure, leading to $\mathbf{Y}\in\mathbb{R}^{B\times T\times C\times D_{Conv}}$.
The receptive field of each \texttt{StackConv} is formulated by:
\begin{equation}
    \mathcal{R}_l = \mathcal{R}_{l-1} + \left( K^{l}-1 \right) \cdot \prod_{i-1}^{l-1}{s^{l}}
    \label{eq:receptive-field}
\end{equation}
where $K^{l}$ and $s^{l}$ are the kernel size and stride at the $l$-th layer, and $\mathcal{R}$ is receptive field with $\mathcal{R}_0=1$.
The multi-scale embedding module aggregates features from $S$ independent \texttt{StackConv} blocks.
Let $\mathbf{Y}_s \in \mathbb{R}^{B \times T \times C \times D_{Conv}}$ denote the output of the $s$-th stack.
\begin{equation}
    \mathbf{x_t} = \mathrm{LayerNorm}\left( 
    \mathrm{GELU}\left( 
    \mathrm{Concat}(\mathbf{Y}_1, \dots, \mathbf{Y}_S) \cdot \mathbf{W}_{\mathrm{emb}}^\top 
    \right) 
  \right), 
  \label{eq:multi_scale_embed}
\end{equation}
where $\mathbf{W}_{\mathrm{emb}} \in \mathbb{R}^{(S\times D_{Conv})\times D_{ti}}$ maps the concatenated features to the target temporal dimension $D_{ti}=512$ for our base model.

Meanwhile, the frequency domain feature is extracted by $\mathbf{x}_{\text{fq}} \in \mathbb{C}^{B \times T \times C \times D_{fq}} = \mathcal{F}(\mathbf{x_p})$, where $\mathcal{F}$ is the FFT operator. 
Here $D_{fq} = \left\lfloor f_s \div 2 \right\rfloor + 1$ due to the Nyquist–Shannon sampling theorem~\cite{jerri1977shannon}, representing the frequency bins in one-sided spectrum.
The single-sided PSD is formulated as $\mathbf{x}_\mathrm{PSD} = |\mathbf{x}_{\text{fq}}|^2$ and normalized as
\begin{equation}
\mathbf{x}_\mathrm{PSD}^{\text{norm}} = \frac{\mathbf{x}_\mathrm{PSD}}{\sum_{p=0}^{P-1} w_{\text{Hann}}^2[p]}
\end{equation}
where the Hanning window is defined as:
\begin{equation}
    w_{\text{Hann}}[p] = 0.5\left(1 - \cos\left(\frac{2\pi p}{P-1}\right)\right)
\end{equation}
where $P$ is patch size. Finally, $\mathbf{x}_\mathrm{PSD}^{\text{norm}}$ is converted to decibels (dB):
\begin{equation}
    \mathbf{x_s} = \log_{10}(\mathbf{x}_\mathrm{PSD}^{\text{norm}}).
\end{equation}
With z-score normalization, the extracted features is formulated as:
\begin{equation}
    \mathbf{x} = \mathrm{Linear}(\mathrm{Z\mathrm{-}Norm}(\mathbf{x_t})) \oplus \mathrm{Linear}(\mathrm{Z\mathrm{-}Norm}(\mathbf{x_s})),
\end{equation}
where $\oplus$ denotes concatenation and the embedding dimension is $D = D_{ti} + D_{fq}$.
The reference sample for reconstruction, combining temporal and frequency domain features, is generated as
\begin{equation}
    \mathbf{x}_{ref} = \mathbf{x_p} \oplus \mathbf{x_s}.
    \label{eq:recon-ref}
\end{equation}

\noindent\textbf{Positional Encoding:}
The channel-wise positional information are encoded via learnable embeddings:
\begin{equation}
    \mathbf{E}_{c} = \mathbf{W}_{pos}[c_1, c_2, \ldots,  c_{|\mathbb{C}|}]^\top \in \mathbb{R}^{|\mathbb{C}| \times d},
\end{equation}
where $c_i$ denotes a channel in the predefined electrode set and the dimension $d$ is given by $d = D/H_q$, where $H_q$ denotes the number of query heads. $\mathbf{W}_{pos}$ is initialized using a truncated normal distribution.
We employ Rotary Positional Encoding~(RoPE)~\cite{su2024roformer} as the temporal relative position encoding $\mathbf{E}_{t}$ to support EEG samples of various lengths.
For query $\mathbf{q}$ and key $\mathbf{k}$ at position $t$:
\begin{equation}
\mathrm{RoPE}(\mathbf{z}_m) = \left[
\begin{aligned}
    & \mathbf{z}_{m,1} \odot \cos(m\theta) - \mathbf{z}_{m,2} \odot \sin(m\theta) \\
    & \mathbf{z}_{m,1} \odot \sin(m\theta) + \mathbf{z}_{m,2} \odot \cos(m\theta)
\end{aligned}
\right]
\end{equation}
where $m$ is the position index, $\mathbf{z}_{m,1}$ denotes the first half of the vector $\mathbf{z}_m$, while $\mathbf{z}_{m,2}$ denotes the second half.
During the forward pass of the model, the sample embeddings $\mathbf{x}$ are augmented by adding $\mathbf{E}_{t}$ along the sequence dimension and $\mathbf{E}_{c}$ along the channel dimension in the channel encoder and decoder, while only $\mathbf{E}_{t}$ is added to the sample embeddings in the temporal encoder.

\subsubsection{\textbf{Channel Encoder}}
Since electrode positions are directly correlated with brain regions, channels play an important role in the analysis of EEG.
However, this property impedes the application of the transformer model to EEG as a type of multivariate data.
The channel encoder unit solves the problem effectively and in parallel by employing a cross-attention mechanism that models inter-channel relationships while preserving temporal patterns.
Built upon an interleaved transformer architecture with rotary positional encoding and learned channel positional encoding, this unit addresses two challenges: 1) Capturing long-range dependencies between non-adjacent EEG channels; and 2) Maintaining temporal coherence during channel-wise feature aggregation.

Given the input tensor $\mathbf{x} \in \mathbb{R}^{B \times T \times C \times D}$ and the channel positional encoding $\mathbf{E}_{c}^\prime$ selected based on the available channels in a specific sample, $\mathbf{x}$ is first reshaped into $\mathbb{R}^{(B \times T) \times C \times D}$ and then the self-attention layer calculations are performed in the same manner as in Equations~\ref{eq:self-attn}~and~\ref{eq:ffn}, which will be discussed in Section~\ref{sec:latent-transformer}.

Then the output representation $\mathbf{h}$ is fed to the cross-attention layer.
We implement cross-attention by reshaping $\mathbf{x}^{\prime} \in \mathbb{R}^{B \times (T \times C) \times D}$, then applying RMS Normalization, linear projection, and RoPE:
\begin{equation}
    Q=W_q \cdot \mathbf{l}_{c}, \ \  K=W_k \cdot \mathrm{RMSNorm}(\mathbf{x}^{\prime}), \ \  V=W_v \cdot \mathrm{RMSNorm}(\mathbf{x}^{\prime}),
\end{equation}
\begin{equation}
    Q^\prime=\mathrm{RoPE}(Q), \quad K^\prime=\mathrm{RoPE}(K)+\mathbf{E}_{c}^\prime, \quad V^\prime=V,
\end{equation}
where $\mathbf{l}_{c} \in \mathbb{R}^{B \times T \times D}$ is the learned query for the cross-attention layer.
Afterwards, cross-attention is formulated as:
\begin{align}
    \mathbf{h} = \mathrm{CrossAttn}(Q^\prime, K^\prime,V^\prime)
    =\mathrm{Softmax}\left({Q^\prime K^{\prime\top}\over\sqrt{d_k}}\odot\mathcal{M}_c \right) V^\prime
\end{align}
where $\mathcal{M}_c$ implements the cross-attention mask described in Section~\ref{sec:pt_ft}.
As mentioned above, two distinct attention layers form a channel encoder layer, which can be stacked to enhance the representation capability~\cite{hu2021unit,pagnoni2024byte}.
For each channel encoder layer $l$, we update the representations as follows:
\begin{align}
    \mathbf{h}_{l}=\mathrm{SelfAttn}(\mathbf{h}_{l-1}); \ \  \mathbf{q}_{l} = \mathrm{CrossAttn}(Q\!=\!\mathbf{q}_{l-1}, K\!=\!V\!=\!\mathbf{h}_l),
\end{align}
where $\mathbf{h}_0=\mathbf{x}^{\prime}$ and $\mathbf{q}_0=\mathbf{l}_{c}$.
Compared to the cross-attention layer, self-attention layers perform computations in a different manner: 1) the query is replaced by $\mathbf{h}_l$, 2) the query is augmented with both $\mathbf{E}_{c}^\prime$ and RoPE, and 3) task-specific attention masks, which are designed to enable interactions among channels at the same time step, are applied to learn robust channel representations.

\subsubsection{\textbf{Temporal Encoder}} \label{sec:latent-transformer}
The temporal encoder captures the general EEG features through stacked self-attention layers, which 1) separate channel and temporal feature aggregation; 2) capture multi-scale temporal dependencies owing to the input embedding; and 3) learn robust representations in various pretraining tasks.

For the input feature, $\mathbf{h_c} \in \mathbb{R}^{B \times T \times D}$ produced by the channel encoder, we introduce a task-specific token $\mathbf{t}_\mathrm{DT} \in \mathbb{R}^{D}$ for different datasets and attach it at the head of the input feature to guide the module in leveraging attention in different patterns, where $\mathbf{h} = \mathbf{t}_\mathrm{DT} \oplus \mathbf{h_c}$.
The self-attention is then formulated as
\begin{equation}
    \mathbf{h_{attn}}=\mathrm{SelfAttn}(\mathrm{RMSNorm}(\mathrm{RoPE}(\mathbf{h})))
    \label{eq:self-attn}
\end{equation}
However, we adopt the innovative Gated Feed-Forward layer (GatedFFN) design in~\cite{pagnoni2024byte} which involves two residual connections and a gated feed-forward layer.
The GatedFFN is formulated as
\begin{equation}
    \begin{aligned}
        \mathbf{h}^* = \mathrm{RoPE}(\mathbf{h}) + W_O \mathbf{h_{attn}};& \ \ 
        \mathbf{h_{norm}}=\mathrm{RMSNorm}(\mathbf{h}); \\
        \mathbf{h_{gate}}=\mathrm{SiLU}(W_\mathrm{gate}\mathbf{h_{norm}});& \ \
        \mathbf{h_{act}}=\mathrm{SiLU}(W_\mathrm{in}\mathbf{h_{norm}}) \odot \mathbf{h_{gate}}; \\
        \mathbf{h_{out}}= \mathbf{h}^* + W_\mathrm{out} \mathbf{h_{act}} ,&
    \end{aligned}
    \label{eq:ffn}
\end{equation}
where all $W_{*}$ denotes weights in linear layers.
The GatedFFN implements parameter-efficient feature interaction, enabling dynamic feature selection through multiplicative interactions and enhancing non-linear representation capability. Dual residual pathways create a gradient highway connection for deep layer training.

\subsubsection{\textbf{EEG Decoder}}
To reconstruct the EEG signal from features, we introduce the EEG decoder includes a hybrid attention mechanism to preserve: 1) temporal coherence in signal dynamics, and 2) channel-specific neuro-physiological patterns.

For the input, $\mathbf{h}_e \in \mathbb{R}^{B \times T \times D}$, from the temporal encoder, and the learnable query $\mathbf{l}_d \in \mathbb{R}^{B \times T \times C \times D}$, RoPE is added to $\mathbf{h}_e$, and both RoPE and channel-specific positional encoding are added to $\mathbf{l}_d$.
The positional encoding serves as a reconstruction guide to achieve a better correlation with the original input.

In contrast to the channel encoder, we utilize a cross-attention layer first, followed by a self-attention layer, to construct a decoder layer, which forms a symmetric decoding.
For the $l$-th decoder layer, 
\begin{align}
    \mathbf{q}_{l}^\prime = \mathrm{CrossAttn}(Q=\mathbf{q}_{l-1}, K=V=\mathbf{h}_e); \ \ \mathbf{q}_{l}=\mathrm{SelfAttn}(\mathbf{q}_l^\prime),
\end{align}
where $\mathbf{q}_0 = \mathbf{l}_d$. 
This symmetric design stabilizes encoder-decoder co-training and eliminates the need for an extra stabilization module.

\subsubsection{\textbf{Pretraining Head}}
Given $\mathbf{q}_{L} \in \mathbb{R}^{B \times T\times C \times D}$ as the output of the EEG decoder, we feed it into two linear projectors for dual-domain reconstruction, $\hat{\mathbf{x}} = [\Phi_t(\mathbf{q}_{L});\Phi_f(\mathbf{q}_{L})]$,
where $\Phi_t$ and $\Phi_f$ aim to project the EEG feature to temporal domain $\mathbb{R}^{D\!} \!\rightarrow\! \mathbb{R}^{D_{ti}}$ and frequency domain $\mathbb{R}^{D\!} \!\rightarrow\! \mathbb{R}^{D_{fq}}$ on feature dimension respectively.

\subsubsection{\textbf{Finetuning Head}}
The finetuning head is designed to implement multi-task adaptive learning on different downstream tasks.
Therefore, the finetuning head is composed of a cross-attention layer and a task-specific \texttt{CLS} token dictionary.
For the EEG feature $\mathbf{h}_e$ from the temporal encoder, the \texttt{CLS} token $\mathbf{t}_\texttt{CLS}^{k}$ for the $k$-th task is used as the query to extract task-specific features from the cross-attention layer, which is formulated as:
\begin{align}
    \mathbf{f}^{k}_\texttt{CLS} = \mathrm{CrossAttn}(Q=\mathbf{t}_\texttt{CLS}^{k}, K=V=\mathbf{h}_e),
\end{align}
where the task-specific features are then projected into the corresponding probability space, $p_\theta(c|\mathbf{t}_{\texttt{CLS}}^{k})$, where $\theta$ is the parameters of the cross-attention layer and the task-specific projector.

\subsection{Training Objectives}\label{sec:pt_ft}
\subsubsection{\textbf{Pretraining Stage}}
Our pretraining framework aims to learn robust EEG representations through a multi-task learning approach
These three complementary pretraining tasks, including 1) \texttt{GPT} for \textbf{EEG signal forecasting}, 2) \texttt{MAE-TP} for \textbf{temporal masked patch reconstruction}, and 3) \texttt{MAE-CH} for \textbf{channel masked patch reconstruction}, achieve temporal dynamics modeling and cross-domain adaptability.
As illustrated in Figure~\ref{fig:model-architecture}, the data patch sequence is obtained and then fed into the pretraining reconstruction head to compute the auto-regressive loss.

\noindent\textbf{Loss function:}
The losses for pretraining are defined as follows. Typically we adopt $\mathrm{RMSE}$ as reconstruction function. For \texttt{GPT} task:
\begin{equation}
    \mathcal{L}_{\mathrm{GPT}}={1\over BC(T-1)} \sum_{n=1}^{B}\sum_{t=1}^{T-1}\sum_{c=1}^{C}\|\hat{\mathbf{x}}_{n,t+1,c} - \mathbf{x}_{n, t,c}\|_2,
    \label{eq:gpt-loss}
\end{equation}
where $B$ is the batch size, $T$ is the number of time steps, and $C$ is the number of channels.
For two \texttt{MAE} pretraining tasks:
\begin{equation}
    \mathcal{L}_{\mathrm{MAE}}={1\over{|\Omega|}} \sum_{(i,j)\in\Omega}\|\hat{\mathbf{x}}_{i,j} - \mathbf{x}_{i,j}\|_2,
    \label{eq:mae-loss}
\end{equation}
where $\Omega$ denotes the set of masked patches, and $(i,j)$ indicates the index $i$ along the time step axis and $j$ along the channel axis, respectively.
Moreover, we introduce a task token, $\mathbf{t}_\mathrm{DT}$, to guide the model to further adapt to datasets collected in different experimental paradigms. 
The task token categorizes the datasets into several classes, including Emotional Recognition, Motor Imaginary, Motor Execution, Seizure Detection, Artifact Classification, Sleep Staging, Resting, Event-Related Potential~(ERP), Visual Stimulus and Workload Estimation. 
Therefore the task classification loss is:
\begin{equation}
    \mathcal{L}_\mathrm{DT}=-{1\over N}\sum_{n=1}^{N}\sum_{c=1}^{C} y_{n, c} \log(p_\theta(c|\mathbf{t}_{\mathrm{DT}}^n),
    \label{eq:data-task-loss}
\end{equation}
where $N$ is the batch size (equal to $B$), $C$ is the number of dataset classes, $y_{n,c}$ is the one-hot ground truth label, and $p_\theta$ is the prediction probability obtained by applying a $\mathrm{Softmax}$ function to the logits.
Finally, the pretrain loss is formulated as
\begin{equation}
    \mathcal{L}_p = \lambda_1\mathcal{L}_{\mathrm{GPT}}+\lambda_2\mathcal{L}_{\mathrm{MAE_{TP}}}+\lambda_3\mathcal{L}_{\mathrm{MAE_{CH}}} + \lambda_4\mathcal{L}_\mathrm{DT},
    \label{eq:pretrain-loss}
\end{equation}
where $\lambda_i$ is set based on the convergence speed of each loss.

\begin{figure}[t]
  \centering
  \includegraphics[width=\columnwidth]{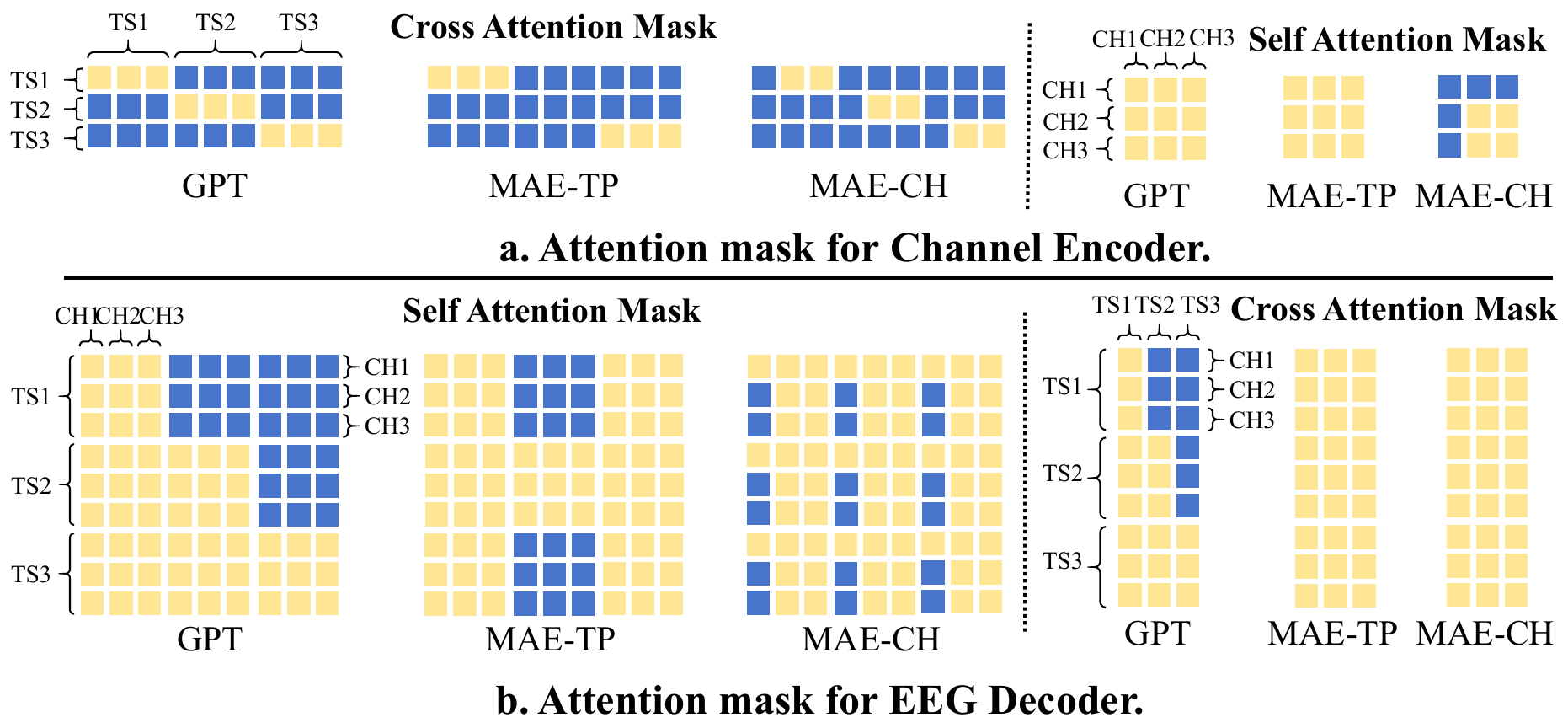}
  \vspace{-10pt}
  \caption{Attention masks in a) Channel Encoder and b) EEG Decoder. Attention is enabled as yellow. Rows and columns correspond to query~(\texttt{Q}) and key~(\texttt{K}) in attention calculation.}
  \label{fig:attn-mask}
  \vspace{-10pt}
\end{figure}

\noindent\textbf{Masking:}
Another important aspect is our masking strategy. 
In supervised objectives, we define three reconstruction losses within an encoder-decoder architecture for EEG representation learning. 
Unlike prior approaches that mask inputs directly or use causal masking in attention, our framework employs diverse masking patterns across modules. 
This strategy separates spatial and temporal operations into dedicated modules and dynamically adjusts masks based on task objectives and training phases, thereby addressing different patch reconstruction needs without modifying the raw input and preventing information leakage. 
For example, when the second time step is masked in \texttt{MAE-TP} and the first channel is masked in \texttt{MAE-CH}, we design an attention mask matrix for each task in the channel encoder as shown in Figure~\ref{fig:attn-mask}~(a). 
In the temporal encoder, corresponding time steps are masked for $\texttt{MAE-TP}$, a lower triangular matrix is used for $\texttt{GPT}$, and an all-ones matrix is applied for $\texttt{MAE-CH}$, while the task token $\mathbf{t}_{\mathrm{DT}}$ remains universally accessible. 
In the decoder, as shown in Figure~\ref{fig:attn-mask}~(b), full sequence masks allow masked patches to attend to others while preventing visible patches from attending to masked ones, reducing noise interference.

\subsubsection{\textbf{Finetuning Stage}}
We design a fine-tuning framework implements multi-task adaptive learning. 
Samples from multiple datasets are fine-tuned on the same foundation model equipped with different classification heads. 
Lightweight classification heads adapt to diverse downstream objectives, while the foundation model enables positive knowledge transfer between related tasks.
In this method, the pretrained representation capabilities are fully exploited, and downstream datasets can be integrated easily and efficiently.
Compared to SOTA multi-task EEG foundation models~\cite{jiang2024neurolm}, there is no need to jointly train LLMs.
As shown in Figure \ref{fig:model-architecture}, we introduce a \texttt{CLS} token $\mathbf{t}_{cls}$ to aggregate information from the compressed representations $\hat{\mathbf{x}}$ for classification.

\noindent\textbf{Loss function:}
During the finetuning stage, $K$ downstream tasks are trained simultaneously. 
The classification loss is formulated as:
\begin{equation}
    \mathcal{L}_{cls}^{k}=-{1\over N_k} \sum_{i=1}^{N_k}\sum_{c=1}^{C_k}w_c^{k}y_{c}^{k}[i]\log \left(p_\theta(c|\mathbf{t}_{\texttt{CLS}}^{k}[i]) \right),
\end{equation}
where $k$ is the task index, $N_k$ is the number of samples in the mini-batch, $C_k$ is the number of classes, $w_c^{k}$ is the class weight computed as inversely proportional to the square root of class frequency, $y_{c}^{k}[i]$ is the one-hot indicator, and $p_\theta(c|\mathbf{t}_{\texttt{CLS}}^{k}[i])$ is the probability that the sample \texttt{CLS} token $\mathbf{t}_{\texttt{CLS}}^{k}[i]$ belongs to class $c$, with $\theta$ denoting the trainable model parameters.
Besides, to preserve the ability of foundation model and prevent catastrophic interference, we introduce a reconstruction loss $\mathcal{L}_{rec}$ calculated as in Equation~\ref{eq:mae-loss} for all input patches.
Finally, the fine-tuning loss is formulated as
\begin{equation}
    \mathcal{L}_f=\sum_{k=1}^{K}{\left( \alpha\mathcal{L}_{cls}^{k}+(1-\alpha)\mathcal{L}_{rec} \right)},
\end{equation}
where $\alpha$ is weight for classification loss. 

\section{Experiments}
\subsection{Datasets}
We follow the dataset composition in~\cite{jiang2024neurolm}, which comprises 15 pretraining datasets and 6 evaluation datasets .
The downstream evaluation dataset includes, \textbf{TUAB}~\cite{7405423}, \textbf{TUEV}~\cite{7405421}, \textbf{TUSL}~\cite{8257018}, \textbf{SEED}~\cite{7104132}, \textbf{HMC}~\cite{10.1371/journal.pone.0256111}, and \textbf{Workload}~\cite{data4010014}.
\textbf{More details about the dataset are in Supplementary Material}.

\subsection{Experimental Setup}
\subsubsection{Data Preprocessing}
An MNE~\cite{10.3389/fnins.2013.00267} pipeline resamples, filters, aligns electrodes, converts units, compresses, and parallelizes EEG data, enabling reproducible large‐scale feature extraction.
\textbf{Detailed data process and evaluation are in Supplementary Material}. 

\subsubsection{Evaluation Metrics}
To address the class imbalance observed in downstream EEG datasets, following the same metrics as in~\cite{jiang2024neurolm}, we adopt Balanced Accuracy, Cohen’s Kappa, and Weighted F1 for multi-category classification, and Balanced Accuracy, AUC-PR, and AUROC for binary classification. \textbf{Detailed descriptions of these metrics are in the Supplementary Material}.

\subsubsection{Baselines}
We compare our framework with three SOTA EEG foundation models~(LaBraM~\cite{jiang2024large}, EEGPT~\cite{wang2024eegpt}, and NeuroLM~\cite{jiang2024neurolm}) and four supervised methods~(SPaRCNet~\cite{jing2023development}, ContraWR~\cite{yang2023self}, FFCL~\cite{li2022motor}, and BIOT~\cite{yang2023biot}), where LaBraM, SPaRCNet, ContraWR, FFCL, and BIOT are designed for single-task scenarios and  NeuroLM and EEGPT are multi-task methods. 
In the experiment, our ALFEE with its three variants (\emph{i.e.}, Medium (M, 44.3M), Base (B, 120M), and Large (L, 300M)), is finetuned in two configurations: (1) a single-task setting, where the pretrained model is finetuned and evaluated on each of the six evaluation tasks separately; and (2) a multi-task setting, where the pretrained model is finetuned and evaluated on all six evaluation tasks jointly. 
\textbf{More details about the baselines, related work, model configurations, and training settings are provided in the Supplementary Material.}

\begin{table*}[ht]
\centering
\caption{Results on TUAB and TUEV. Best performance on multi-task and single-task are highlight in bolded.}
\vspace{-10pt}
\resizebox{\textwidth}{!}{%
\setlength\tabcolsep{14pt}
\begin{tabular}{@{}lccccccc@{}}
\toprule
\multirow{2}{*}{\textbf{Methods}} & \multirow{2}{*}{\textbf{Multi-task}} & \multicolumn{3}{c}{\textbf{TUAB}} & \multicolumn{3}{c}{\textbf{TUEV}} \\
\cmidrule(lr){3-5} \cmidrule(lr){6-8}
& & \textbf{Balanced Acc.} & \textbf{AUC-PR} & \textbf{AUROC} & \textbf{Balanced Acc.} & \textbf{Cohen’s Kappa} & \textbf{Weighted F1} \\
\midrule
SPaRCNet & \ding{55} & 0.7896$\pm$0.0018 & 0.8414$\pm$0.0018 & 0.8676$\pm$0.0012 & 0.4161$\pm$0.0262 & 0.4233$\pm$0.0181 & 0.7024$\pm$0.0104 \\
ContraWR & \ding{55} & 0.7746$\pm$0.0041 & 0.8421$\pm$0.0104 & 0.8456$\pm$0.0074 & 0.4384$\pm$0.0349 & 0.3912$\pm$0.0237 & 0.6893$\pm$0.0136 \\
FFCL & \ding{55} & 0.7848$\pm$0.0038 & 0.8448$\pm$0.0065 & 0.8569$\pm$0.0051 & 0.3979$\pm$0.0104 & 0.3732$\pm$0.0188 & 0.6783$\pm$0.0120 \\
BIOT & \ding{55} & 0.7959$\pm$0.0057 & 0.8792$\pm$0.0023 & 0.8815$\pm$0.0043 & 0.5281$\pm$0.0225 & 0.5273$\pm$0.0249 & 0.7492$\pm$0.0082 \\
LaBrAM-base & \ding{55} & 0.8140$\pm$0.0019 & 0.8965$\pm$0.0016 & \textbf{0.9022$\pm$0.0009} & 0.6409$\pm$0.0065 & 0.6637$\pm$0.0093 & 0.8312$\pm$0.0052 \\
\midrule
ALFEE-M    & \ding{55}  & 0.8069$\pm$0.0109 & 0.8835$\pm$0.0191 & 0.8828$\pm$0.0142 & 0.6190$\pm$0.0261 & 0.6612$\pm$0.0299 & 0.7953$\pm$0.0132 \\
ALFEE-B    & \ding{55}  & 0.8108$\pm$0.0072 & 0.8899$\pm$0.0043 & 0.8859$\pm$0.0019 & 0.6439$\pm$0.0052 & 0.7529$\pm$0.0073 & 0.8481$\pm$0.0093 \\
ALFEE-L    & \ding{55}  & \textbf{0.8239$\pm$0.0091} & \textbf{0.9042$\pm$0.0059} & 0.9015$\pm$0.0028 & \textbf{0.6525$\pm$0.0049} & \textbf{0.7737$\pm$0.0103} & \textbf{0.8626$\pm$0.0124} \\
\toprule
EEGPT      & \checkmark & 0.7983$\pm$0.0030 & -                 & 0.8718$\pm$0.0050 & 0.6232$\pm$0.0114 & 0.6351$\pm$0.0134 & 0.8187$\pm$0.0063 \\
NeuroLM-B & \checkmark  & 0.7826$\pm$0.0065 & 0.6975$\pm$0.0081 & 0.7816$\pm$0.0079 & 0.4560$\pm$0.0048 & 0.4285$\pm$0.0048 & 0.7153$\pm$0.0028 \\
\midrule
ALFEE-M    & \checkmark & 0.7951$\pm$0.0048 & 0.8515$\pm$0.0162 & 0.8589$\pm$0.0039 & 0.6553$\pm$0.0124 & 0.6607$\pm$0.0098 & 0.7959$\pm$0.0052 \\
ALFEE-B    & \checkmark & 0.8074$\pm$0.0082 & 0.8588$\pm$0.0171 & 0.8763$\pm$0.0091 & \textbf{0.7173$\pm$0.0102} & 0.7254$\pm$0.0222 & 0.8342$\pm$0.0093 \\
ALFEE-L    & \checkmark & \textbf{0.8090$\pm$0.0072} & \textbf{0.8861$\pm$0.0252} & \textbf{0.8812$\pm$0.0082} & 0.6987$\pm$0.0145 & \textbf{0.7863$\pm$0.0273} & \textbf{0.8683$\pm$0.0102} \\
\bottomrule
\end{tabular}}
\label{tab:results1}
\vspace{-5pt}
\end{table*}

\begin{table*}[ht]
\centering
\caption{Results on SEED and HMC. Results of EEGPT are reproduced based on public released code.}
\vspace{-10pt}
\resizebox{\textwidth}{!}{%
\setlength\tabcolsep{14pt}
\begin{tabular}{@{}lccccccc@{}}
\toprule
\multirow{2}{*}{\textbf{Methods}} & \multirow{2}{*}{\textbf{Multi-task}} & \multicolumn{3}{c}{\textbf{SEED}} & \multicolumn{3}{c}{\textbf{HMC}} \\
\cmidrule(lr){3-5} \cmidrule(lr){6-8}
& & \textbf{Balanced Acc.} & \textbf{Cohen's Kappa} & \textbf{Weighted F1} & \textbf{Balanced Acc.} & \textbf{Cohen’s Kappa} & \textbf{Weighted F1} \\
\midrule
SPaRCNet        & \ding{55} & 0.5596$\pm$0.0244 & 0.3464$\pm$0.0372 & 0.5585$\pm$0.0297 & 0.4756$\pm$0.1109 & 0.3147$\pm$0.1315 & 0.4108$\pm$0.1310 \\
ContraWR        & \ding{55} & 0.6106$\pm$0.0078 & 0.4220$\pm$0.0129 & 0.6137$\pm$0.0085 & 0.4242$\pm$0.0541 & 0.2340$\pm$0.0554 & 0.2987$\pm$0.0288 \\
FFCL            & \ding{55} & 0.5808$\pm$0.0322 & 0.3732$\pm$0.0462 & 0.5743$\pm$0.0402 & 0.4427$\pm$0.0702 & 0.2542$\pm$0.0654 & 0.2902$\pm$0.0485 \\
BIOT            & \ding{55} & 0.7097$\pm$0.0024 & 0.5682$\pm$0.0051 & 0.7134$\pm$0.0027 & 0.6862$\pm$0.0041 & 0.6295$\pm$0.0113 & 0.7091$\pm$0.0147 \\
LaBrAM-base     & \ding{55} & \textbf{0.7318$\pm$0.0019} & \textbf{0.5994$\pm$0.0031} & \textbf{0.7354$\pm$0.0021} & 0.7286$\pm$0.0101 & 0.6812$\pm$0.0073 & 0.7554$\pm$0.0024 \\
\midrule
ALFEE-M         & \ding{55}  & 0.6561$\pm$0.0088 & 0.4854$\pm$0.0049 & 0.6572$\pm$0.0093 & 0.7229$\pm$0.0105 & 0.6636$\pm$0.0091 & 0.7378$\pm$0.0066 \\
ALFEE-B         & \ding{55}  & 0.6963$\pm$0.0029 & 0.5461$\pm$0.0021 & 0.6924$\pm$0.0038 & \textbf{0.7490$\pm$0.0133} & 0.6922$\pm$0.0109 & \textbf{0.7627$\pm$0.0092} \\
ALFEE-L         & \ding{55}  & 0.6931$\pm$0.0035 & 0.5414$\pm$0.0052 & 0.6848$\pm$0.0045 & 0.7338$\pm$0.0162 & \textbf{0.7083$\pm$0.0134} & 0.7720$\pm$0.0112 \\
\toprule
EEGPT      & \checkmark & 0.7122$\pm$0.0022 & 0.5734$\pm$0.0049 & 0.7099$\pm$0.0038 & 0.7029$\pm$0.0082 & 0.6584$\pm$0.0059 & 0.7323$\pm$0.0041 \\
NeuroLM-B  & \checkmark & 0.5554$\pm$0.0075 & 0.3393$\pm$0.0117 & 0.5599$\pm$0.0068 & 0.6737$\pm$0.0050 & 0.6188$\pm$0.0057 & 0.7126$\pm$0.0034 \\
\midrule
ALFEE-M    & \checkmark & 0.6578$\pm$0.0034 & 0.4880$\pm$0.0062 & 0.6592$\pm$0.0082 & 0.7197$\pm$0.0073 & 0.6782$\pm$0.0083 & 0.7551$\pm$0.0039 \\
ALFEE-B    & \checkmark & 0.7248$\pm$0.0052 & 0.5887$\pm$0.0089 & 0.7221$\pm$0.0102 & \textbf{0.7388$\pm$0.0098} & 0.6815$\pm$0.0078 & 0.7584$\pm$0.0082 \\
ALFEE-L    & \checkmark & \textbf{0.7411$\pm$0.0072} & \textbf{0.6153$\pm$0.0078} & \textbf{0.7432$\pm$0.0079} & 0.7378$\pm$0.0074 & \textbf{0.6837$\pm$0.0068} & \textbf{0.7610$\pm$0.0083} \\
\bottomrule
\end{tabular}}
\label{tab:results2}
\vspace{-5pt}
\end{table*}

\begin{table*}[ht]
\centering
\caption{Results on Workload and TUSL. Results of EEGPT are reproduced based on public released code.}
\vspace{-10pt}
\resizebox{\textwidth}{!}{%
\setlength\tabcolsep{14pt}
\begin{tabular}{@{}lccccccc@{}}
\toprule
\multirow{2}{*}{\textbf{Methods}} & \multirow{2}{*}{\textbf{Multi-task}} & \multicolumn{3}{c}{\textbf{Workload}} & \multicolumn{3}{c}{\textbf{TUSL}} \\
\cmidrule(lr){3-5} \cmidrule(lr){6-8}
& & \textbf{Balanced Acc.} & \textbf{AUC-PR} & \textbf{AUROC} & \textbf{Balanced Acc.} & \textbf{Cohen’s Kappa} & \textbf{Weighted F1} \\
\midrule    
SPaRCNet        & \ding{55} & 0.5977$\pm$0.0071 & 0.6638$\pm$0.0314 & 0.6717$\pm$0.0172 & 0.4185$\pm$0.0452 & 0.1399$\pm$0.0799 & 0.3500$\pm$0.0968 \\
ContraWR        & \ding{55} & 0.6966$\pm$0.0332 & 0.7668$\pm$0.0408 & 0.7685$\pm$0.0317 & 0.5857$\pm$0.0662 & 0.3567$\pm$0.0968 & 0.5458$\pm$0.0798 \\
FFCL            & \ding{55} & 0.7069$\pm$0.0197 & \textbf{0.7823$\pm$0.0099} & 0.7857$\pm$0.0234 & 0.3819$\pm$0.0688 & 0.0628$\pm$0.0888 & 0.2120$\pm$0.0786 \\
BIOT            & \ding{55} & 0.6655$\pm$0.0665 & 0.7189$\pm$0.0722 & 0.7342$\pm$0.0536 & 0.5758$\pm$0.0303 & 0.2012$\pm$0.0212 & 0.2394$\pm$0.0040 \\
LaBrAM-base     & \ding{55} & 0.6609$\pm$0.0204 & 0.7174$\pm$0.0234 & 0.7272$\pm$0.0165 & 0.7625$\pm$0.0131 & 0.6407$\pm$0.0304 & \textbf{0.7614$\pm$0.0210} \\
\midrule
ALFEE-M         & \ding{55}  & 0.6409$\pm$0.0298 & 0.3315$\pm$0.0272 & 0.7661$\pm$0.0379 & 0.7262$\pm$0.0201 & 0.5217$\pm$0.0322 & 0.6588$\pm$0.0208 \\
ALFEE-B         & \ding{55}  & 0.6518$\pm$0.0183 & 0.5922$\pm$0.0211 & 0.8004$\pm$0.0138 & 0.7535$\pm$0.0182 & \textbf{0.6417$\pm$0.0277} & 0.7573$\pm$0.0299 \\
ALFEE-L         & \ding{55}  & \textbf{0.7333$\pm$0.0201} & 0.6420$\pm$0.0232 & \textbf{0.8151$\pm$0.0172} & \textbf{0.7698$\pm$0.0129} & 0.6077$\pm$0.0329 & 0.7336$\pm$0.0310 \\
\toprule
EEGPT      & \checkmark & 0.6299$\pm$0.0178 & 0.6792$\pm$0.0092 & 0.6928$\pm$0.0108 & 0.7288$\pm$0.0143 & 0.5972$\pm$0.0209 & 0.7233$\pm$0.0153 \\
NeuroLM-XL & \checkmark & 0.6172$\pm$0.0113 & 0.5824$\pm$0.0080 & 0.6253$\pm$0.0160 & 0.6734$\pm$0.0436 & 0.5107$\pm$0.0617 & 0.6743$\pm$0.0394 \\
\midrule
ALFEE-M    & \checkmark & 0.6285$\pm$0.0223 & 0.4268$\pm$0.0272 & 0.6339$\pm$0.0249 & 0.7292$\pm$0.0111 & 0.6283$\pm$0.0219 & 0.7490$\pm$0.0182 \\
ALFEE-B    & \checkmark & 0.6849$\pm$0.0239 & \textbf{0.6871$\pm$0.0282} & \textbf{0.8295$\pm$0.0292} & 0.8245$\pm$0.0172 & 0.7441$\pm$0.0282 & 0.8312$\pm$0.0143 \\
ALFEE-L    & \checkmark & \textbf{0.7333$\pm$0.0244} & 0.5516$\pm$0.0258 & 0.7992$\pm$0.0283 & \textbf{0.8680$\pm$0.0181} & \textbf{0.8158$\pm$0.0302} & \textbf{0.8799$\pm$0.0154} \\
\bottomrule
\end{tabular}}
\label{tab:results3}
\vspace{-10pt}
\end{table*}

\begin{figure*}[t]
  \centering
  \includegraphics[width=\textwidth]{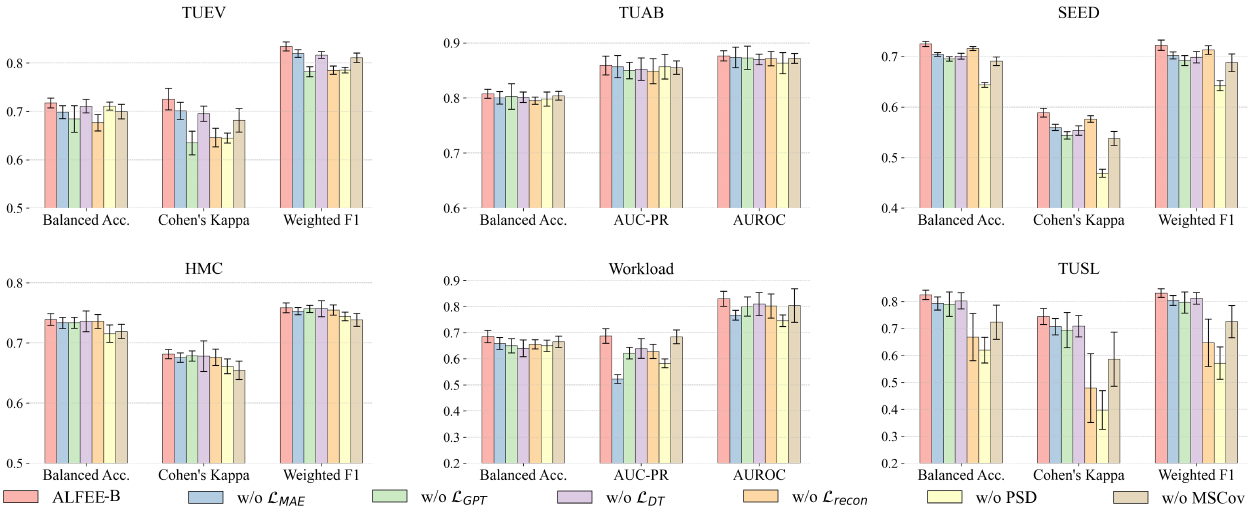}
  \vspace{-10pt}
  \caption{Ablation Study. We study the effect of four loss functions~($\mathrm{w/o}$ $\mathcal{L}_{\mathrm{MAE}}$, $\mathrm{w/o}$ $\mathcal{L}_{\mathrm{GPT}}$, and $\mathrm{w/o}$ $\mathcal{L}_{\mathrm{DT}}$, $\mathrm{w/o}$ $\mathcal{L}_{recon}$), and two feature extraction module~(the frequency domain feature, $\mathrm{w/o}$ $\mathrm{PSD}$ and the multi-scale convolution, $\mathrm{w/o}$ $\mathrm{MSCov}$).}
  \label{fig:ablation}
  \vspace{-10pt}
\end{figure*}

\subsection{Main Results}

In Tables~\ref{tab:results1},~\ref{tab:results2}, and~\ref{tab:results3}, we present all the experimental results on both single-task and multi-task configuration. 
The best results on both settings are highlighed in bolded.
All experiment are conducted for ten times, with mean $\pm$ standard deviation reported.
\textbf{Scaling law experiments are in Supplementary Material.}

\noindent\textbf{Single-task Performance:}
In our single-task configuration, we evaluate ALFEE-M, ALFEE-B and ALFEE-L on each downstream dataset independently and compare their performance with several baselines, most notably LaBrAM-base.
Overall, ALFEE-B achieves results comparable to LaBrAM-base on most datasets, while exhibiting higher Balanced Accuracy, Cohen’s Kappa, and Weighted F1 scores in TUEV and HMC.
These results underscore ALFEE-B’s ability to capture essential EEG features with a relatively moderate model size.
When we scale up to ALFEE-L, the performance improves further on most tasks, indicating that increasing model capacity confers distinct advantages in extracting nuanced EEG patterns.
For instance, ALFEE-L exhibits superior sensitivity and specificity on TUAB, a dataset that contains more complex signals for medical diagnosis.
Although LaBrAM-base remains competitive on certain metrics (\emph{e.g.}, on SEED), our results show that ALFEE-B and ALFEE-L surpass it in most single-task settings.
Beyond the numerical improvements, our analysis reveals two key characteristics of ALFEE in the single-task setting. 
First, the unified architecture of ALFEE-B achieves stable training dynamics and converges efficiently, suggesting that it can accommodate differences in input distributions
Second, as the model size varies from ALFEE-M to ALFEE-L, the larger model capacity of ALFEE-L allows it to capture subtle temporal and channel variations, leading to improved generalization on tasks requiring complex pattern recognition
However, this increased capacity also introduce overfitting risk, particularly in datasets with fewer samples (\emph{e.g.}, Workload). 
Overall, our single-task evaluations confirm that ALFEE provides competitive and often superior performance compared to existing baselines.

\noindent\textbf{Multi-task Performance:}
In the multi-task configuration, we evaluate ALFEE across four model sizes, ALFEE-M, ALFEE-B, ALFEE-L, on six datasets. 
Compared with NeuroLM and EEGPT, our models consistently achieve higher performance across all metrics.
Specifically, on the SEED dataset, ALFEE-M attains a significant improvement (10\%–15\% across three metrics) over NeuroLM, while ALFEE-L pushes the performance even further (20\%–26\% across three metrics), surpassing single-task approaches and providing evidence of multi-task synergy.
Notably, on TUEV, ALFEE-L exceeds the performance of LaBrAM, reaffirming that a well-designed multi-task approach can outperform single-task strategies
A similar trend is also observed on HMC, where ALFEE-B and ALFEE-L demonstrate clear advantages over the baselines and maintain a robust margin of improvement in all metrics.
Furthermore, on TUSL, despite its limited data size, our larger models still outperformed EEGPT and NeuroLM, emphasizing the flexibility of ALFEE in handling small-scale datasets without severe overfitting.
One of the key observations is that scaling up our model size from ALFEE-M to ALFEE-L tends to achieves better performance in most tasks. 
At the same time, our results on tasks with fewer samples suggest that heavier models require careful early stopping strategies to avoid overfitting.
The complementary information shared across tasks in the multi-task setting contributes to the improved performance, enabling ALFEE to acquire more robust representations of EEG signals.
This effect is particularly meaningful because it provides a more efficient and unified framework for training, obviating the need for intensive task-specific fine-tuning.
Overall, ALFEE outperform multi-task methods, demonstrating that a larger model capacity is beneficial for representation learning with appropriately dataset balance.

\subsection{Ablation Study}
In this section, we study the effect of loss functions~(w/o $\mathcal{L}_{\mathrm{MAE}}$, w/o $\mathcal{L}_{\mathrm{GPT}}$, and w/o $\mathcal{L}_{\mathrm{DT}}$ for the pretraining stage, and w/o $\mathcal{L}_{recon}$ for the finetuning stage) and two feature extraction modules (the frequency domain feature, w/o PSD, and the multi-scale convolution, w/o MSCov) using ALFEE-B as shown in Figure~\ref{fig:ablation}.

\noindent\textbf{Effect of Different Losses:} 
Comparing the performance among the three pretraining losses, we notice that $\mathcal{L}_\mathrm{GPT}$ and $\mathcal{L}_\mathrm{MAE}$ have a relatively large impact on downstream performance; without them, performance would degrade by approximately 2\% to 10\% across different metrics.
Moreover, $\mathcal{L}_\mathrm{GPT}$ also has a large impact on model stability, as the standard deviation for ``w/o $\mathcal{L}_\mathrm{GPT}$'' in TUAB, TUEV, HMC, and TUSL is two times larger than that of ALFEE-B.
Considering that temporal forecasting is a more advanced ability than mask reconstruction, such instability also indicates the effectiveness of $\mathcal{L}_\mathrm{GPT}$ for robust EEG representation.
As for $\mathcal{L}_{\mathrm{DT}}$, it has little impact on overall performance but is helpful for model stability because it aids the model in capturing task-specific characteristics beyond channel relationships and temporal dynamics during pretraining.
In the finetuning stage, $\mathcal{L}_\mathrm{recon}$ is very useful, particularly when the downstream dataset is small, such as TUSL and Workload, thereby proving its effectiveness in preventing catastrophic interference during multi-task finetuning.

\noindent\textbf{Effect of Different Feature Extraction Modules:}
For the feature extraction modules, we notice that both are very important for downstream performance.
The multi-scale convolution captures temporal dynamics at different granularities, thereby providing additional temporal patterns for downstream classification, especially in TUSL, HMC, and SEED datasets. 
The frequency domain feature is critical for TUEV, SEED, Workload, and TUSL datasets.
For the Workload dataset, the $\gamma$-wave in the frequency domain is very important for attention assessment.
For the TUSL dataset, the $\beta$-wave in the frequency domain helps detect seizures.
For SEED and TUEV, the electromyographic signal in the frequency domain from the face can introduce noise during EEG collection, therefore, frequency domain feature can filter out such noise.
These results suggest the effectiveness of our loss functions and feature extractors in the ALFEE for robust EEG representation.

\section{Conclusion}

In this paper, we propose the ALFEE framework, a EEG foundation model that incorporates a hybrid attention architecture to separate channel-wise feature aggregation from temporal dynamics modeling, enabling robust EEG representation with variable channel configurations.
By separating EEG representations into adaptive channel-wise and task-guided temporal components, ALFEE leverages a hybrid attention architecture and a dual-stage optimization process, employing multi-task pretraining followed by multi-task fine-tuning with cross-attention layers, to reconstruct both temporal and frequency domain features and address the diverse challenges inherent in EEG signal representation.
Our extensive experiments across six diverse EEG datasets demonstrate the superior performance of the ALFEE framework in EEG signal representation and multi-task learning.
Overall, ALFEE represents a major advancement in brain-computer interfaces and healthcare, underscoring the significant potential of EEG foundation models in signal processing and multi-task learning.
Moreover, scaling law analysis confirms that large-scale models such as ALFEE offer promising prospects for EEG applications and human-machine interactions.

\appendix
In the Appendix, we provide related work in Appendix~\ref{sec:related} to introduce the progress in the relevant research field.
We also provide detailed descriptions of the datasets used in our two learning stages in Appendix~\ref{sec:data}, and experimental settings in Appendix~\ref{sec:exp_setting}.
Moreover, we conduct an additional experiment to validate the scaling law in EEG signal representation in Appendix~\ref{sec:scaling}.
Finally, we conclude by discussing the limitations and future work in Appendix~\ref{sec:limit}.

\section{Related Work}~\label{sec:related}
\subsection{EEG Modeling Approaches}
Traditional EEG analysis initially relies on time-frequency features extracted through the discrete wavelet transform~\cite{chen2017high} or PSD~\cite{alsolamy2016emotion} via Fourier transforms, combined with classifiers such as support vector machines or multilayer perceptrons.
While these methods achieve reasonable performance in brain information decoding, their limited generalizability across experimental paradigms and subjects prompts the development of deep learning architectures.
Early convolutional neural networks, including ConvNet~\cite{schirrmeister2017deep} and EEGNet~\cite{lawhern2018eegnet}, establish end-to-end pipelines through spatiotemporal convolutions, though their capacity to capture long-range dependencies remained constrained.
To address this limitation, recurrent neural networks, such as ChronoNet~\cite{subhrajit2019chrono} and ATDD-LSTM~\cite{du2020efficient}, are introduced, thereby improving temporal modeling at the cost of training efficiency and gradient stability.

Graph neural networks like EEG-GNN~\cite{demir2021eeg} subsequently emerge, leveraging the electrode-channel relationships for neuroscientifically interpretable analysis.
However, their fixed node representations often compromise temporal resolution and adaptability.
The recent shift toward attention mechanisms and transformer architectures has yielded notable advancements in EEG analysis.
ST-Transformer~\cite{song2021transformer} pioneers transformer-based EEG analysis, while EEGConformer~\cite{song2022eeg} hybridizes convolutional local feature extraction with global self-attention.
Despite their potential, these models face challenges in cross-paradigm generalization and often require extensive datasets to mitigate overfitting.
Self-supervised learning approaches have further addressed data scarcity. 
REMoNet~\cite{jiang2024remonet} employs channel masking and spectral feature prediction for emotion recognition, while EEG2Rep~\cite{foumani2024eeg2rep} learns noise-invariant representations through abstract feature reconstruction.
However, their reliance on task-specific feature designs limits broader applicability across diverse experimental paradigms.

\begin{table*}[h]
    \caption{Detailed information about pretraining datasets.}
    \label{tab:pretrain-dataset}
    \centering
    \begin{tabular}{ccccccm{6.5cm}}
    \toprule
    Dataset &  Category & \#Channel & Duration & \#Train & \#Valid & \makecell[c]{Description}\\[1ex]
    \midrule
    TUEG~\cite{obeid2016temple} & \makecell{Clinical\\Recordings}  &  60 & 60 & 1515391 & 75436 & A rich archive of 26,846 clinical EEG recordings collected at Temple University Hospital. \\ [1ex]
    \hline
    SEED-IV~\cite{8283814} & \makecell{Emotion\\Recognition}& 60 & 10 & 13678 & 977 & A emotion EEG dataset that 15 subjects watch 72 film clips which have the tendency to induce happiness, sadness, fear or neutral emotions.\\[1ex]
    \hline
    SEED-V~\cite{9395500} & \makecell{Emotion\\Recognition} & 60 & 10 & 13246 & 2088 & A multi-modal emotion dataset that 20 subjects watch video clips in happy, sad, disgust, fear and neutral. \\[1ex]
    \hline
    SEED-GER~\cite{liu2022identifying} & \makecell{Emotiom\\Recognition} & 60 & 10 & 8440 & 919 & A emotion dataset eight German subjects watch 20 film clips (positive, neutral, negative) as stimuli. \\[1ex]
    \hline
    SEED-FRA~\cite{liu2022identifying}  & \makecell{Emotion\\Recognition} & 60 & 10 & 6846 & 978 & Eight French subjects watch 21 film clips in French (positive, neutral, negative) as stimuli. \\[1ex]
    \hline
    BCIC-1A~\cite{BLANKERTZ2007539} & \makecell{Motor\\Imagery} & 43 & 8 & 3155 & 535 & EEG recordings for motor imagery tasks, where subjects imagined moving either their left hand, right hand, or foot. \\[1ex]
    \hline
    Emobrain~\cite{savran2006emobrain} & \makecell{Emotion\\Recognition} & 54 & 10 & 1370 & 405 & Multimodal emotion detection dataset using brain signals (EEG, fNIRS) from 5 male subjects. \\[1ex]
    \hline
    Grasp and Lift~\cite{Luciw2014} & \makecell{Motor\\Execution} & 32 & 5 & 7003 & 1390 & Grasp and lift action from 12 subjects in total,  10 series of trials for each subject.\\[1ex]
    \hline
    Inria BCI P300~\cite{Margaux2012} & ERP & 56 & 5 & 13647 & 7901 & A P300-based spelling dataset including 26 subjects. \\[1ex]
    \hline
    \makecell{Motor\\Movement\\Imagery~\cite{1300799}} & \makecell{Motor\\Imagery} & 64 & 12 & 13516 & 650 & 1500 EEG recordings dataset obtained from 109 volunteers perform opens four action in both in imaginary and reality. \\[1ex]
    \hline
    \makecell{Resting State\\(Trujillo 2017)~\cite{10.3389/fnins.2017.00425}} & Resting & 64 & 10 & 981 & 100 & A dataset comprising 22 subjects for a resting eyes closed and eyes open. \\[1ex]
    \hline
    \makecell{Raw EEG Data\\(Trujillo 2019)~\cite{10.3389/fnins.2019.01292}} & \makecell{Visual\\Stimulus} & 64 & 30 & 4806 & 319 & EEG was recorded during reported Information-Integration categorization and reported multidimensional Rule-Based categorization tasks. \\[1ex]
    \hline
    \makecell{Siena Scalp\\EEG Database~\cite{pr8070846}} & \makecell{Seizure\\Detection} & 27 & 40 & 8260 & 4408 & EEG recordings of 14 patients that are labeled epilepsy and the classification of seizures is carefully reviewed.\\[1ex]
    \hline
    \makecell{SPIS Resting\\State~\cite{9034192}} & Resting & 64 & 10 & 270 & 30 & A resting-state EEG from 10 subjects contains 2.5 minutes of eyes-open and 2.5 minutes of eyes-closed. \\[1ex]
    \hline
    \makecell{Target Versus\\Non-Target~\cite{korczowski:hal-02172347}} & ERP & 32 & 15 & 3403 & 333 & dataset contains EEG recordings of 50 subjects playing to a visual P300 Brain-Computer Interface (BCI) video game. \\[1ex]
    \bottomrule
    \end{tabular}
\end{table*}

\begin{table*}[ht]
    \caption{Detailed information about evaluation datasets.}
    \label{tab:finetune-dataset}
    \centering
    \begin{tabular}{cccccccc}
    \toprule
    Dataset &  Category & \#Channel & Duration & \#Train & \#Valid & \#Test & Task \\[1ex]
    \midrule
    TUAB      & Clinical Recording & 23    & 30 & 247728 & 12315 & 12277 &  Binary Classification  \\[1ex]
    TUEV      & Artifact Detection & 21    & 5 & 87834  & 12473 & 13046 &  6-class Classification \\[1ex]
    TUSL      & Seizure Detection & 21,22 & 10 & 222    & 43    & 25    &  3-class Classification \\[1ex]
    SEED      & Emotion Recognition & 60    & 10 & 22455  & 7875  & 7560  &  3-class Classification \\[1ex]
    HMC       & Sleep Staging & 4     & 30 & 91681  & 22804 & 22440 &  5-class Classification \\[1ex]
    Workload  & Workload Estimation & 19    & 10 & 1537   & 300   & 297   &  Binary Classification  \\[1ex]
    \bottomrule
    \end{tabular}
\end{table*}

\subsection{Transformer in Foundation Models}
The impact of self-supervised transformers in natural language processing~(NLP), as exemplified by the masked language modeling of BERT~\cite{devlin2019bert} and the autoregressive pretraining of GPT~\cite{radford2018improving,radford2019language,mann2020language,achiam2023gpt}, has catalyzed cross-modal adaptations.
Vision transformers~\cite{dosovitskiy2020image} demonstrate that transformer architectures can surpass convolutional networks in image classification by processing sequences of image patches.
CLIP~\cite{radford2021learning} extends this paradigm through contrastive learning on massive image-text pairs, achieving unprecedented zero-shot generalization.
MAE~\cite{he2022masked} further advances latent representation learning by reconstructing randomly masked image patches with an asymmetric encoder-decoder design.
Recently, BLT~\cite{pagnoni2024byte} introduces a novel architecture with alternating self-attention and cross-attention layers, enabling dynamic attention allocation across variable-length sequences, which is also a key inspiration for our hybrid attention mechanism.

While transformers dominate NLP and computer vision, their adoption in time series analysis remains nascent. 
For univariate forecasting, Lag-Llama~\cite{rasul2023lag} establishes a foundation model using autoregressive training on diverse datasets, demonstrating robust generalization to unseen domains.
Multivariate approaches face greater complexity: PatchTST~\cite{nie2022time} addresses this through channel-independent patch embedding strategies, significantly improving long-term forecasting accuracy.
Crossformer~\cite{zhang2023crossformer} enhances cross-variate dependency modeling via dimension-aware attention, while Pathformer~\cite{chen2024pathformer} integrates multi-scale temporal resolutions through adaptive pathway selection.
Morai~\cite{woo2024unified} introduces innovations like multiple patch size projections and any-variate attention to handle heterogeneous sampling rates and variable input dimensions.
Despite these advances, existing models primarily target structured numerical data (\emph{e.g.}, stock prices, statistical data), lacking mechanisms to handle the unique challenges of bioelectrical signals—non-stationarity, low signal-to-noise ratios, and inter-subject variability—which limit their direct applicability to EEG analysis.

\subsection{Multi-Task Learning}
The scarcity of labeled data for individual tasks motivates the adoption of multi-task learning to enhance data utilization efficiency by mining inter-task relationships and learning shared representations, thereby improving model generalization and transferability.
MT-DNN~\cite{liu2019multi} demonstrates robust domain adaptation capabilities by integrating four distinct task categories, also effectively mitigating overfitting while promoting universal feature learning.
T5~\cite{raffel2020exploring} standardizes diverse NLP tasks into a unified text-to-text framework, revealing that multi-task pretraining with subsequent fine-tuning achieves performance parity with single-task pretraining approaches. 
UniLM~\cite{dong2019unified} extends the architecture of BERT~\cite{devlin2019bert} through dynamic attention masking mechanisms, enabling a single model architecture to address heterogeneous language tasks without structural modifications.
OmniNet~\cite{pramanik2019omninet} employs HogWild parallel training to facilitate cross-modal knowledge transfer in multi-task scenarios, while UniT~\cite{hu2021unit} implements an end-to-end unified transformer that concurrently learns seven distinct tasks across eight datasets through joint training, achieving competitive performance while being parameter-efficient.
For temporal data processing, UniTS~\cite{gao2024building} develops a domain-agnostic framework by integrating multi-domain datasets, demonstrating the superiority of specialized time series transformers over language-oriented LLMs in handling classification, forecasting, and imputation tasks.
Nevertheless, these existing architectures require substantial adaptation to effectively process time-series neurophysiological data and achieve optimal performance in EEG applications.

\subsection{Large Scale Pretrained EEG Models}
The success of self-supervised learning in language modeling has spurred its adoption for EEG analysis, particularly given the scarcity of clinical annotations. 
BENDR~\cite{kostas2021bendr} pioneers self-supervision to learn general EEG data distributions, while EEG2Vec~\cite{bethge2022eeg2vec} introduces a conditional variational autoencoder framework for joint generative-discriminative representation learning.
BIOT~\cite{yang2023biot} addresses practical challenges such as variable electrode configurations and signal durations, demonstrating enhanced performance on clinical benchmarks.
BrainBERT~\cite{wang2023brainbert} adapts the methodology of BERT~\cite{devlin2019bert} to stereo-electroencephalography analysis through time-frequency representation learning.
The emergence of large-scale pretrained EEG transformer models has significantly enhanced generalization capacity and analytical precision in biosignal processing.
The Brant~\cite{zhang2023brant,yuan2024brant,zhang2024brant} series models, trained on terabyte-scale datasets, exhibit exceptional robustness against data variability and excellent scalability, even demonstrating proficiency in modeling EEG-physiological signal correlations.
LaBraM~\cite{jiang2024large} currently achieves SOTA performance through its innovative integration of VQ-VAE~\cite{van2017neural} modules with dual-domain (frequency/phase) autoregressive learning.
NeuroLM~\cite{jiang2024neurolm} bridges neurosignal-language modality gaps by embedding EEG signal into pretrained LLM frameworks, thereby balanced performance across unified downstream tasks.
EEGPT~\cite{wang2024eegpt} advances the field through the combination of dual self-supervised universal representation learning and stabilization mechanisms inspired by MoCo~\cite{he2020momentum}.
However, challenges such as variant channel counts, inadequate channel-temporal supervision, and upstream–downstream domain gaps still limit the performance of existing methods.

\section{Datasets Description}\label{sec:data}

The detailed information about all 15 pretraining datasets are listed in Table \ref{tab:pretrain-dataset}. All data in the duration column is measured in seconds.
Besides, the detailed information about the evaluation datasets is listed in Table \ref{tab:finetune-dataset}: 1) The \textbf{TUAB}~\cite{7405423} dataset contains EEG records that are classified as clinically normal or abnormal; 2) The \textbf{TUEV}~\cite{7405421} dataset contains sessions that include events such as periodic lateralized epileptiform discharges, generalized periodic epileptiform discharges, spike and/or sharp wave discharges, artifacts, and eye movements; 3) The \textbf{TUSL}~\cite{8257018} dataset contains sessions that include seizure events, slowing events, and complex background events; 4) The \textbf{SEED}~\cite{7104132} dataset contains physiological signal data and corresponding emotion labels. Fifteen Chinese movie clips were carefully selected as stimuli to evoke different emotions; 5) The \textbf{HMC}~\cite{10.1371/journal.pone.0256111} dataset collects 151 whole-night polysomnographic (PSG) sleep recordings as well as event annotations corresponding to the scoring of sleep patterns (hypnogram) performed by sleep technicians; 6) The \textbf{Workload}~\cite{data4010014} dataset contains EEG recordings from 36 healthy volunteers during mental serial subtraction along with corresponding reference background EEGs. Based on task performance, subjects are divided into two groups: one for background and one for arithmetic.

\section{Experimental Setting}
\label{sec:exp_setting}

\subsection{Data Preprocessing}
The dataset preprocessing pipeline employs a structured approach for handling EEG data from various datasets using MNE-tools~\cite{10.3389/fnins.2013.00267}.
The pipeline initializes with a resampling operation that transforms the source sampling rate to $f_s=256 ,\text{Hz}$, which facilitates patch division.
An overlap-add Finite Impulse Response~(FIR) high-pass filter is applied to remove low-frequency noise, otherwise the signal length is too short to meet the required filter length.
Subsequently, a $50,\text{Hz}$ or $60,\text{Hz}$ notch filter is applied after human review.
Afterwards, electrode configurations from the datasets are aligned with the predefined 10-10 channel set.
Data unit conversion is performed from $\mathrm{\mu V}$ to Volts for MNE compatibility.
For specific implementation, EEG data is serialized in Parquet format with Zstandard compression to expedite dataset loading. Remote storage is supported via the S3 protocol for distributed computing.
This processing pipeline enables reproducible feature extraction while maintaining physiological signal fidelity.
The implementation leverages parallel processing for efficient large-scale data handling.

\subsection{Baselines}
We mainly consider three SOTA EEG foundation models, namely LaBraM~\cite{jiang2024large}, EEGPT~\cite{wang2024eegpt}, and NeuroLM~\cite{jiang2024neurolm}, as our baseline methods. 
LaBraM is pre-trained on 2,500 hours of data with integrated VQ-VAE~\cite{van2017neural} modules for dual-domain (frequency/phase) mask learning, EEGPT combines dual self-supervised universal representation learning and stabilization mechanisms, and NeuroLM bridges neurosignal-language modality gaps by embedding EEG signals into pretrained LLM frameworks using 25,000 hours of data.
In addition, we select four supervised methods for comparison, including SPaRCNet~\cite{jing2023development}, ContraWR~\cite{yang2023self}, FFCL~\cite{li2022motor}, and BIOT~\cite{yang2023biot}
Among these methods, LaBraM, SPaRCNet, ContraWR, FFCL, and BIOT are designed for single-task scenarios and are also used as baselines in NeuroLM~\cite{jiang2024neurolm}. Therefore, we adopt the reported results from NeuroLM~\cite{jiang2024neurolm} in our experiments.
Besides, since NeuroLM~\cite{jiang2024neurolm} and EEGPT~\cite{wang2024eegpt} are multi-task methods, we also compare with them under multi-task settings.
Since EEGPT~\cite{wang2024eegpt} does not report its performance on the SEED, HMC, Workload, and TUSL datasets, we reproduce the results using its official, publicly available pretrained model weights and code.
We follow the default settings in their linear\_prob algorithm for training on these datasets.

\subsection{Evaluation Strategy}
We implement a greedy algorithm-based multi-label stratified splitting function to divide a dataset into training, validation, and test sets by subject, while ensuring that the label distribution in each split is balanced and aligns with predefined ratios.
Subsequently, dataset partitioning is performed by splitting subjects into training, validation, and test sets as follows: 1) \textbf{TUAB}: The validation and test sets are obtained by equally splitting the original evaluation set by subject, while the training set remains unchanged; 2) \textbf{TUEV} and \textbf{TUSL}: Owing to the highly imbalanced label distribution in these datasets, the stratified splitting function is employed to create three splits from all the data, approximately aligning with predefined ratios of 0.8, 0.1, and 0.1; 3) \textbf{SEED}: Following prior research, the 15 trials are divided into three sets in a 9:3:3 ratio, and all sessions are merged together thereafter; 4) \textbf{HMC}: Subjects are randomly split into training, validation, and test sets at a ratio of 103:24:24; 5) \textbf{Workload}: Stratified splitting is employed to achieve approximate ratios of 0.72, 0.14, and 0.14 for training, validation, and test sets, respectively.

\subsection{Evaluation Metrics}
To address the class imbalance commonly observed in downstream EEG datasets, the following evaluation metrics are adopted for performance comparison:
\begin{itemize}
    \item \textbf{Balanced Accuracy}: the arithmetic mean of recall (sensitivity) across all classes, mitigating the impact of imbalanced class distributions. It is particularly effective for evaluating classification models on datasets with significant disparities in class proportions.
    \item \textbf{Weighted F1}: a harmonic mean of precision and recall, weighted by the number of true instances in each class. This metric accounts for class imbalance by assigning higher importance to classes with larger sample sizes, ensuring a more representative evaluation of model effectiveness.
    \item \textbf{AUROC}: area under the ROC curve. It reflects the model’s ability to discriminate between classes across all possible decision boundaries.
    \item \textbf{AUC-PR}: area under the precision-recall curve. It provides a holistic evaluation of model performance under class imbalance.
    \item \textbf{Cohen’s Kappa}: the agreement level between predicted and true labels by comparing observed and expected frequencies along the diagonal of a confusion matrix. It is particularly suited for multi-class classification scenarios.
\end{itemize}
Among these metrics, AUROC and AUC-PR are used to evaluate binary classification tasks, while Cohen's Kappa and Weighted F1 are applied to multi-category classification. 
Together, these metrics provide a robust evaluation framework under class imbalance.

\subsection{Training Settings}
To facilitate data loading, all samples in the datasets are transformed into an Arrow dataset after cleaning and preprocessing, thereby speeding up distributed computing and leveraging the GPU’s direct data access functionality.
All experiments are conducted using Python 3.11.11, PyTorch 2.6.0, and CUDA 12.4 on eight A100 GPUs for pretraining and two A800 GPUs for finetuning.
We enable autonomous mixed precision in the bfloat16 data type to improve GPU memory utilization and introduce \texttt{GradScaler} to prevent gradient explosion.
We employ the AdamW optimizer and a two-phase learning rate scheduler, which combines linear warmup with cosine annealing.
During the pretraining stage, the learning rate is set to 1$\times$10$^{-4}$, and the hyperparameters $\lambda_1$, $\lambda_2$, $\lambda_3$, and $\lambda_4$ are set to 0.4, 0.275, 0.275, and 0.05, respectively, to balance the different loss terms.
During the finetuning stage, the learning rate is set to 5$\times$10$^{-5}$, and $\alpha$ is set to 0.9 to prevent catastrophic interference.
To compare with existing methods under different settings, the finetuning stage is conducted in two configurations: (1) a single-task setting, where the pretrained model is finetuned and evaluated on each of the six evaluation tasks separately; and (2) a multi-task setting, where the pretrained model is finetuned and evaluated on all six evaluation tasks jointly.
The best models are trained on the training set, selected based on performance on the validation set, and finally evaluated on the test set.

\subsection{Model Configurations}\label{sec:model_config}
ALFEE has five variants: Small~(S), Medium~(M), Base~(B), Large~(L) and Extra Large~(XL), with 16.3M, 44.3 M, 120 M, 300 M, and 540 M parameters, respectively. 
We can alter the parameter scale by changing the number of stacked transformer blocks and the embedding dimensionality. 
The maximum learnable query length and sequence length are set to 2048 to accommodate the longest input EEG patch sequences. 
The patch size is equal to $f_s=256$, representing 1 second.
50\% of channels and 40\% of time steps are randomly masked. 
More details regarding the settings of the model variants are listed in Table \ref{tab:conf-variant}.

\begin{table*}[ht]
    \caption{Configurations for model variants.}
    \label{tab:conf-variant}
    \centering
    \begin{tabular}{@{}lccccc@{}}
    \toprule
     & ALFEE-S & ALFEE-M & ALFEE-B & ALFEE-L & ALFEE-XL \\
    \midrule
    \#Parameters & 16.3 M & 44.3 M & 120 M & 300 M & 540 M \\[1ex]
    Model Dim & 384 & 512 & 640 & 896 & 1152 \\[1ex]
    MLP Dim & 256 & 512 & 512 & 768 & 768 \\[1ex]
    \makecell[l]{Channel Block} & 1 & 1 & 2 & 2 & 3 \\[1ex]
    \makecell[l]{Temporal Block} & 5 & 7 & 14 & 16 & 19 \\[1ex]
    \makecell[l]{Decoder Block} & 2 & 4 & 8 & 10 & 12 \\[1ex]
    Batch Size & 256 & 256 & 128 & 96 & 64 \\[1ex]
    Attn. Head & 4 & 4 & 8 & 8 & 12 \\[1ex]
    \bottomrule
    \end{tabular}
\end{table*}

\begin{figure*}[!ht]
  \centering
  \begin{subfigure}[t]{0.3\textwidth}
    \includegraphics[width=\linewidth]{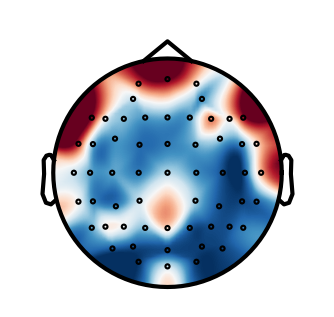}
    \caption{Sad}
    \label{fig:grad-cam-sad}
  \end{subfigure}
  \hfill
  \begin{subfigure}[t]{0.3\textwidth}
    \includegraphics[width=\linewidth]{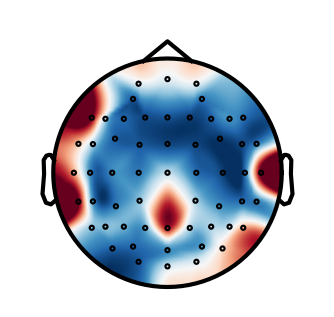}
    \caption{Neutral}
    \label{fig:grad-cam-neutral}
  \end{subfigure}
  \hfill
  \begin{subfigure}[t]{0.3\textwidth}
    \includegraphics[width=\linewidth]{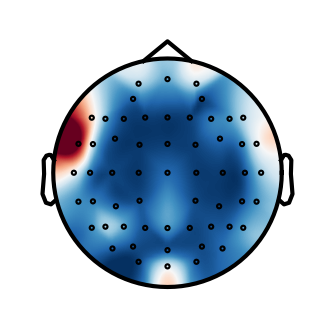}
    \caption{Happy}
    \label{fig:grad-cam-happy}
  \end{subfigure}
  \caption{Average Grad-CAM visualization showing the model's region of interest for target class prediction. Warmer colors indicating higher relevance, generated by computing gradient flow of channel encoder on ALFEE-B.}
  \label{fig:grad-cam}
  \Description{Visualization of Grad-CAM for sad emotion.}
\end{figure*}

\begin{figure*}[!ht]
    \centering
    \begin{subfigure}[t]{0.48\textwidth} 
        \includegraphics[width=\linewidth]{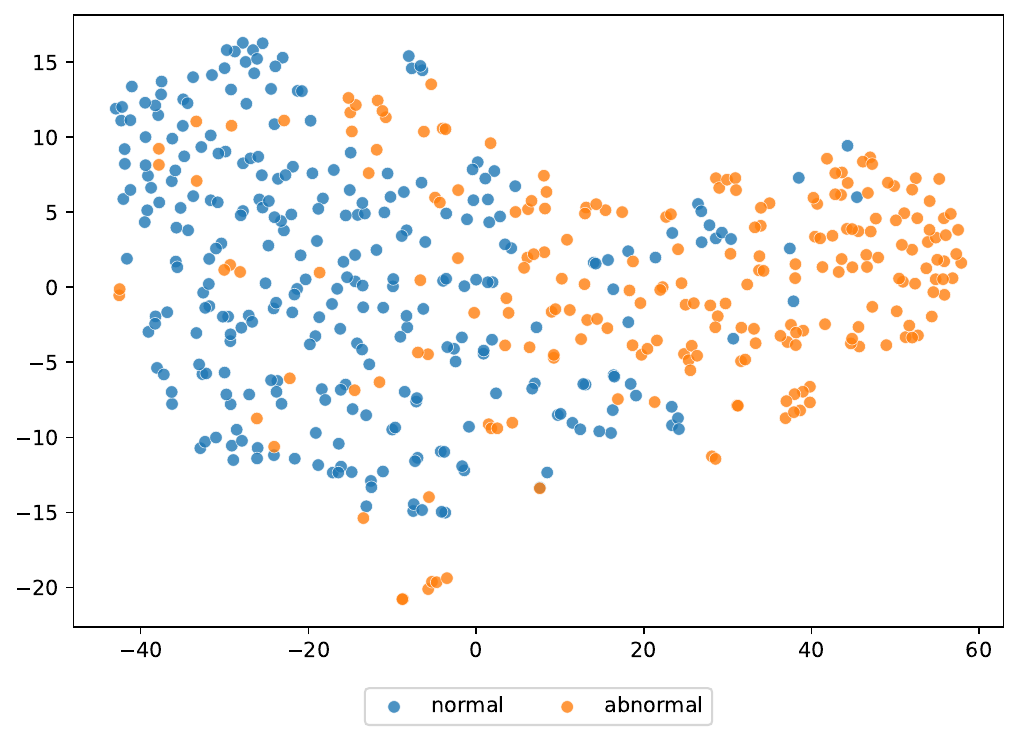}
        \caption{TUAB}
        \label{fig:t-sne-tuab}
    \end{subfigure}
    \hfill
    \begin{subfigure}[t]{0.48\textwidth}
        \includegraphics[width=\linewidth]{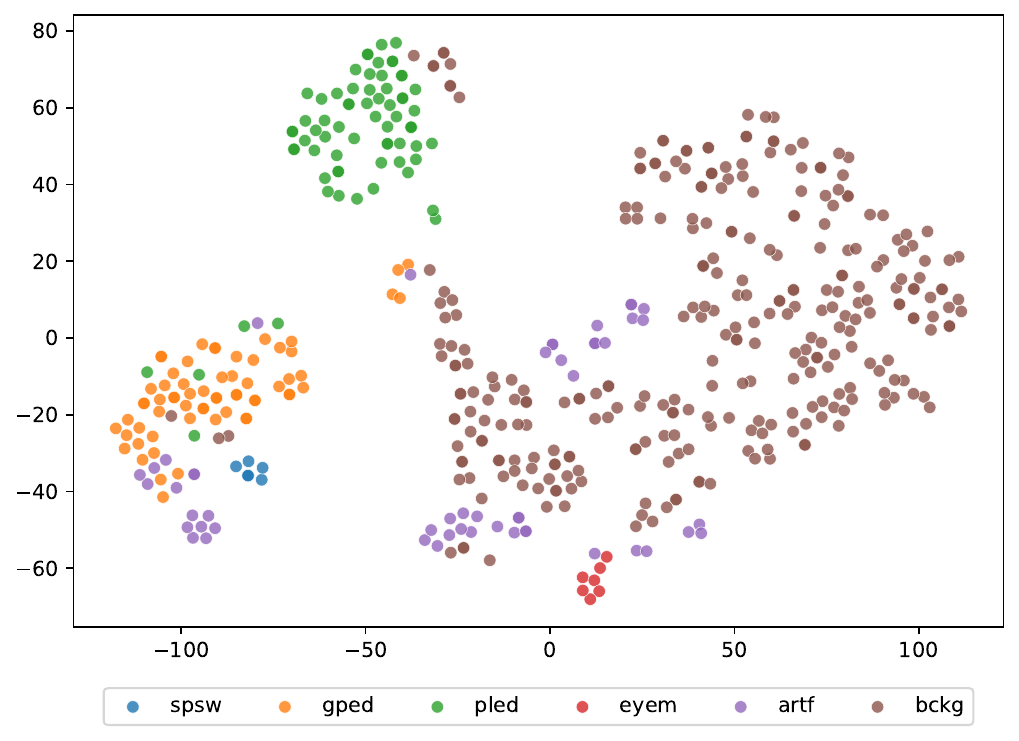}
        \caption{TUEV}
        \label{fig:t-sne-tuev}
    \end{subfigure}
    \vspace{1em}
    \begin{subfigure}[t]{0.48\textwidth}
        \includegraphics[width=\linewidth]{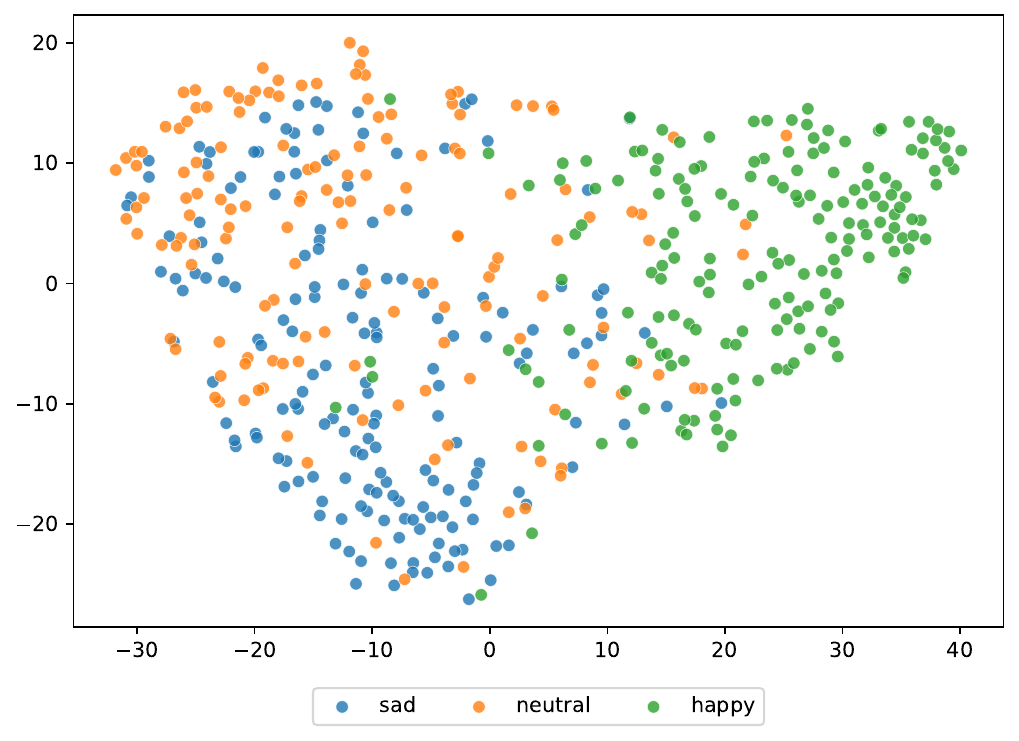}
        \caption{SEED}
        \label{fig:t-sne-seed}
    \end{subfigure}
    \hfill
    \begin{subfigure}[t]{0.48\textwidth}
        \includegraphics[width=\linewidth]{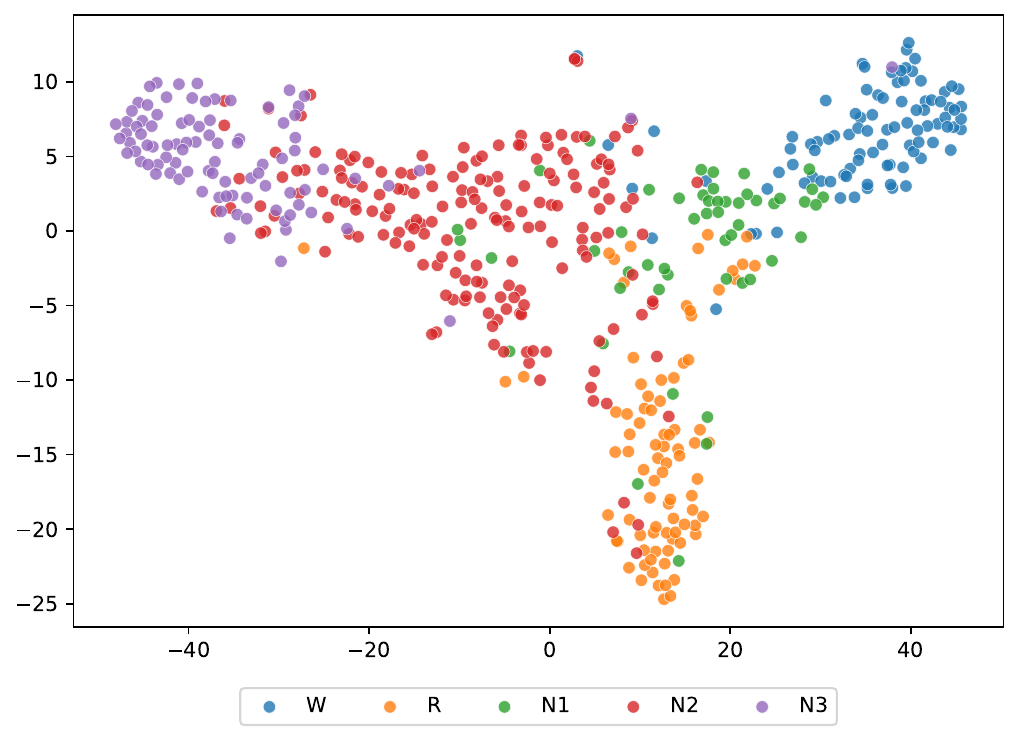}
        \caption{HMC}
        \label{fig:t-sne-hmc}
    \end{subfigure}
    \caption{t-SNE visualizations of feature embeddings reduced to 40 dimensions by PCA on different datasets with ALFEE-B. t-SNE runs for 1500 iterations in which the perplexity is 30 and the number of samples is 500.}
    \label{fig:t-sne}
    \Description{Visualization of t-SNE classification cluster result.}
\end{figure*}

\section{Visualization}
\label{sec:visual}

\subsection{Interpretable Decision Localization}
In this section, we exploit Gradient-weighted Class Activation Mapping~(Grad-CAM)~\cite{selvaraju2017grad} on the SEED dataset to reveal how our ALFEE framework captures emotional information from the spatial channels of EEG signals.
Specifically, Grad-CAM is a visualization technique that highlights critical regions in input images that influence CNN predictions.
By computing gradient-based weights from the target class score to the original EEG features, the resulting heatmaps provide intuitive explanations for model decisions.
This technique has also been adopted to extract information about activated brain regions from model parameters and to perform visualization to enhance explainability in neuroscience~\cite{liu2023fine}.
In our implementation, we attach it to the channel encoder to visualize channel-wise decision-making evidence.

The attention of our ALFEE framework to different target classes may vary across individuals; however, the averaged Grad-CAM visualization on a standard brain model reflects the model’s overall attention patterns.
As shown in Figure~\ref{fig:grad-cam}, this visualization confirms that our ALFEE framework focuses on relevant electrodes or brain regions for classification.
Apparent lateralization can be observed, which is consistent with previous research on the functional networks of emotion~\cite{gainotti2019role}.
In Figure~\ref{fig:grad-cam-sad}, our framework pays more attention to Fp1, Fpz, F4, F8, and T8 electrodes, suggesting that the right hemisphere may be dominant in negative emotion.
For positive emotion, the left frontal lobe around F7 and FT7 receives strong attention from our framework.
Additionally, other regions including Pz, TP7, and Oz can be viewed as the basis for classification in neutral emotion.
All of these results are consistent with established scientific evidence~\cite{liu2023fine,pessoa2017network}.

\subsection{Latent Feature Clustering via t-SNE}
t-SNE is a nonlinear dimensionality reduction technique designed to visualize high-dimensional data in a low-dimensional space (typically 2D or 3D) while preserving local structures and cluster relationships~\cite{vandermaaten08a}.
In Figure~\ref{fig:t-sne}, we visualize the features of six datasets from the classification head of ALFEE to demonstrate the representational ability of our framework.

For the TUAB dataset, as shown in Figure~\ref{fig:t-sne-tuab}, two lateral clusters emerge with a broad transitional region, which are approximately linearly separable.
For the TUEV dataset, as shown in Figure~\ref{fig:t-sne-tuev}, the six predefined classes show strong separation, except for the diffuse artifact category spanning the embedding space, consistent with its miscellaneous definition during the annotation process.
For the SEED dataset, as shown in Figure~\ref{fig:t-sne-seed}, positive and negative emotions form distinct clusters, while neutral samples exhibit intermediate positioning with partial overlap, likely due to inter-individual variability in perception.
For the HMC dataset, as shown in Figure~\ref{fig:t-sne-hmc}, the five sleep stages form tightly grouped clusters, reflecting high intra-class consistency aligned with physiological patterns.
These results confirm the representation ability of ALFEE to disentangle discriminative features while tolerating noise.

\section{Scaling Laws}
\label{sec:scaling}

In this section, we study the relationship between model performance and training scale, including the model size and pretraining data size.
For the model size, we design 5 variants, as described in Appendix~\ref{sec:model_config}, to investigate the effects of model size on pretraining loss and downstream task accuracy.
For the data size, we change the pretraining data from 2,500 hours to 25,000 hours on ALFEE-B to investigate the effects of pretraining dataset on pretraining loss and downstream task accuracy.
The pretraining loss is calculated on a small partition of validation pretraining data, and the downstream task accuracy is calculated based on the balanced accuracy on SEED and TUSL dataset.

\subsection{For Parameter Size}
\begin{figure*}[!ht]
  \centering
  \begin{subfigure}[t]{0.3\textwidth}
    \includegraphics[width=\linewidth]{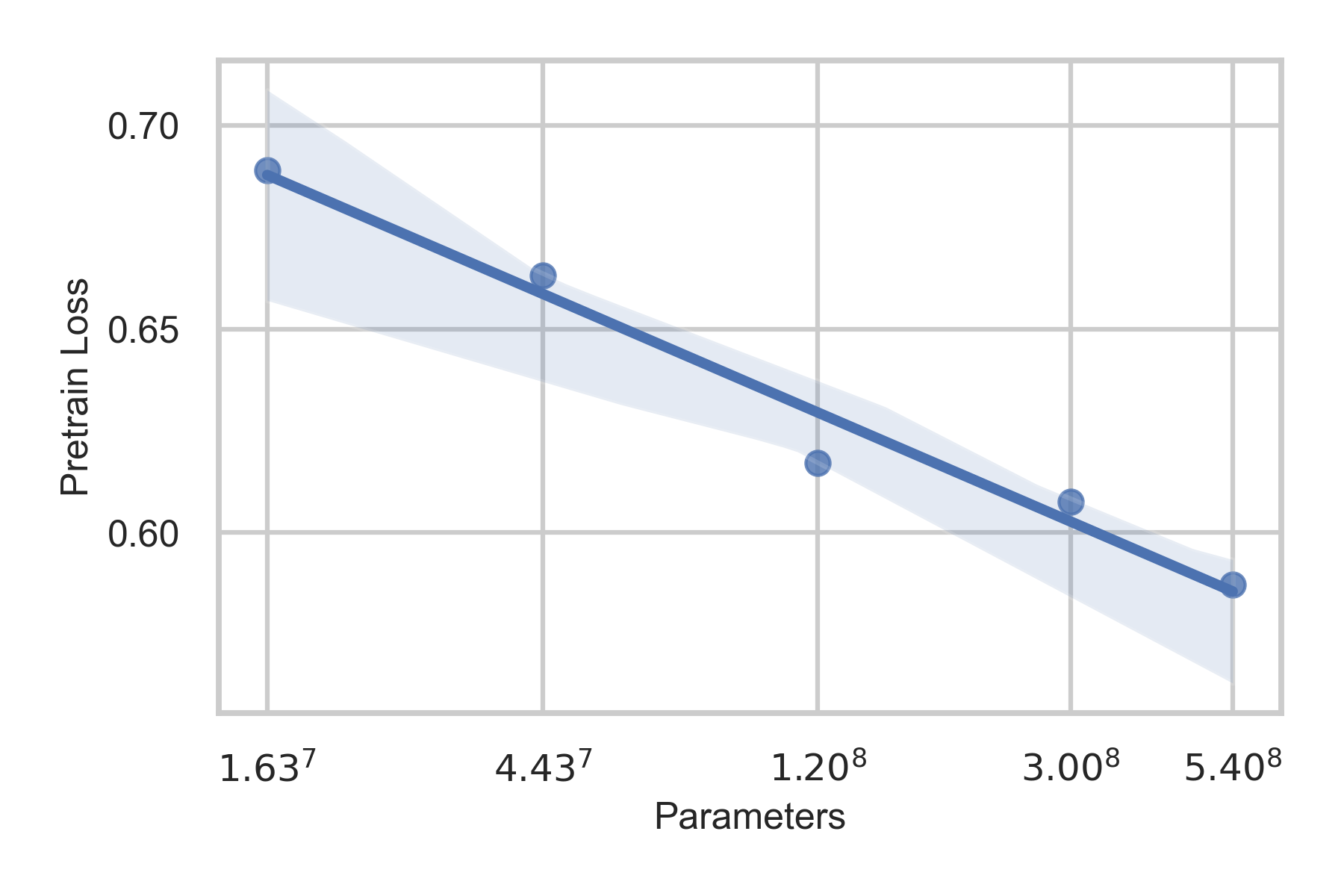}
    \caption{Scaling laws with model size $N$ and $\mathcal{L}_p$}
    \label{fig:param-loss}
  \end{subfigure}
  \hfill
  \begin{subfigure}[t]{0.3\textwidth}
    \includegraphics[width=\linewidth]{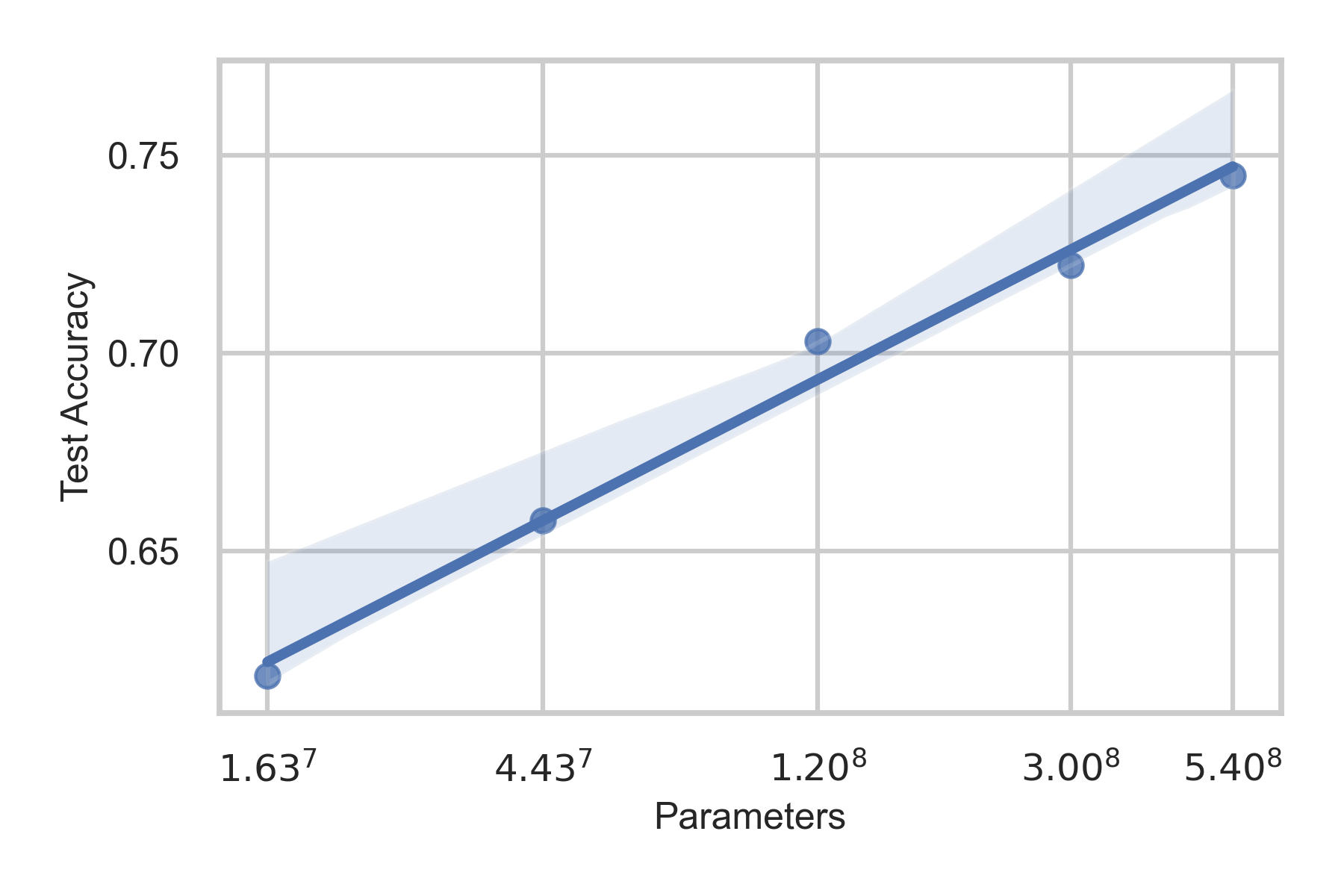}
    \caption{Scaling laws with model size $N$ and SEED's balanced accuracy}
    \label{fig:param-seed}
  \end{subfigure}
  \hfill
  \begin{subfigure}[t]{0.3\textwidth}
    \includegraphics[width=\linewidth]{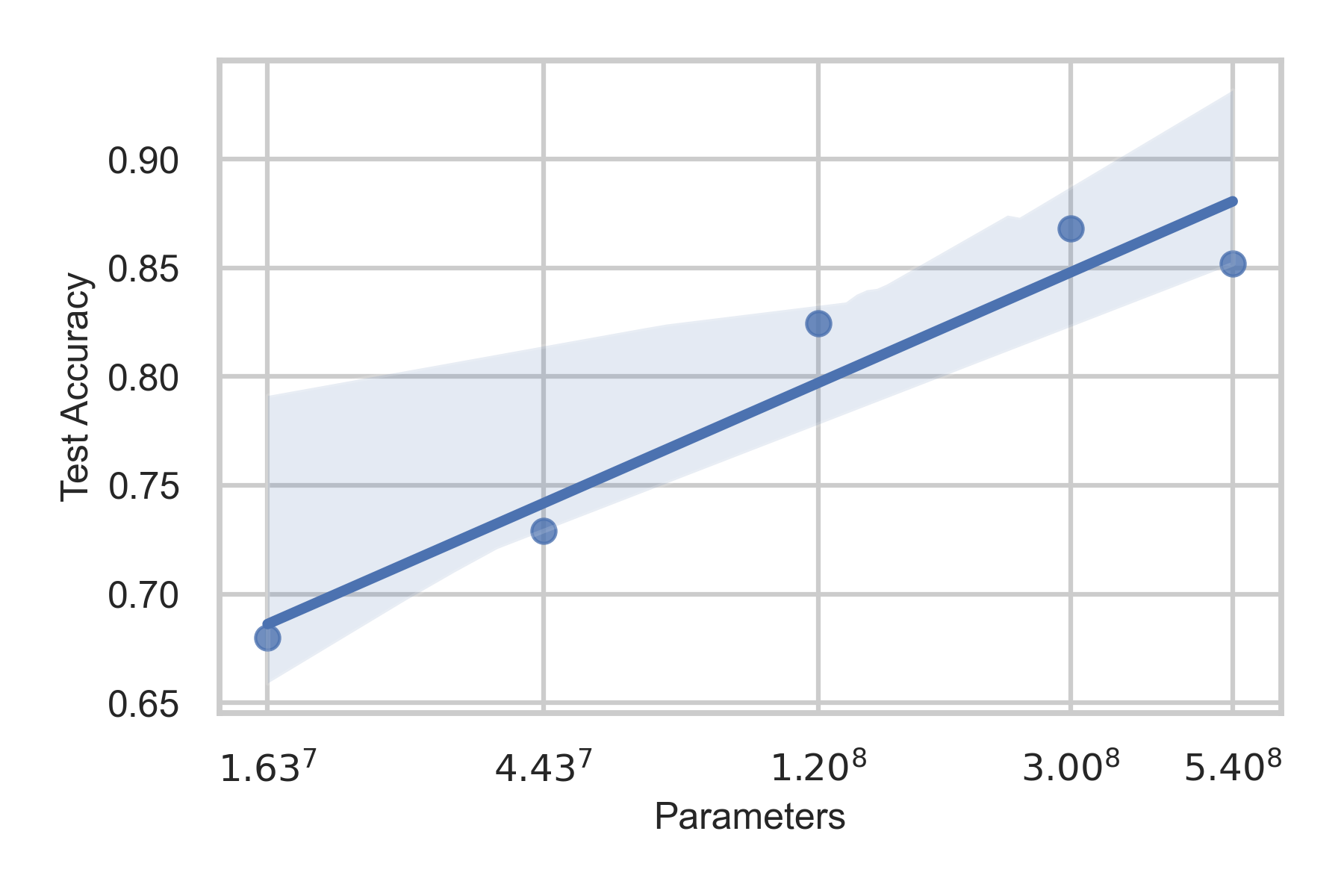}
    \caption{Scaling laws with model size $N$ and TUSL's balanced accuracy}
    \label{fig:param-tu}
  \end{subfigure}
  \caption{Scaling laws with model size $N$ and a) $\mathcal{L}_p$; b) SEED balanced accuracy; c) TUSL balanced accuracy. Axes are all on a logarithmic scale.}
  \label{fig:param}
\end{figure*}

In Figures~\ref{fig:param-loss},~\ref{fig:param-seed},~\ref{fig:param-tu}, we provide the results of the scale law experiments on the pretraining loss function~($\mathcal{L}_p$), SEED, and TUSL dataset.
The results on the pretraining loss function show that the scaling law of the test $\mathcal{L}_p$ with model size~($N$) is: \color{black}$\mathcal{L}_p$ = $-0.029$ * $\ln$(N) + $1.173$ \color{black}, where $R^2$ is $0.972$.
The results on the SEED dataset show that the scaling law of the test balanced accuracy with model size~($N$) is: \color{black}$BAcc$ = $0.036$ * $\ln$(N) + $0.028$\color{black}, where $R^2$ is $0.988$.
The results on the TUSL dataset show that the scaling law of the test balanced accuracy with model size~($N$) is: \color{black}$BAcc$ = $0.056$ * $\ln$(N) - $0.236$\color{black}, where $R^2$ is $0.919$.
The results of the pretraining loss and downstream task balanced accuracy indicate that larger models generally achieve higher accuracy. 

\begin{figure*}[!ht]
  \centering
  \begin{subfigure}[t]{0.3\textwidth}
    \includegraphics[width=\linewidth]{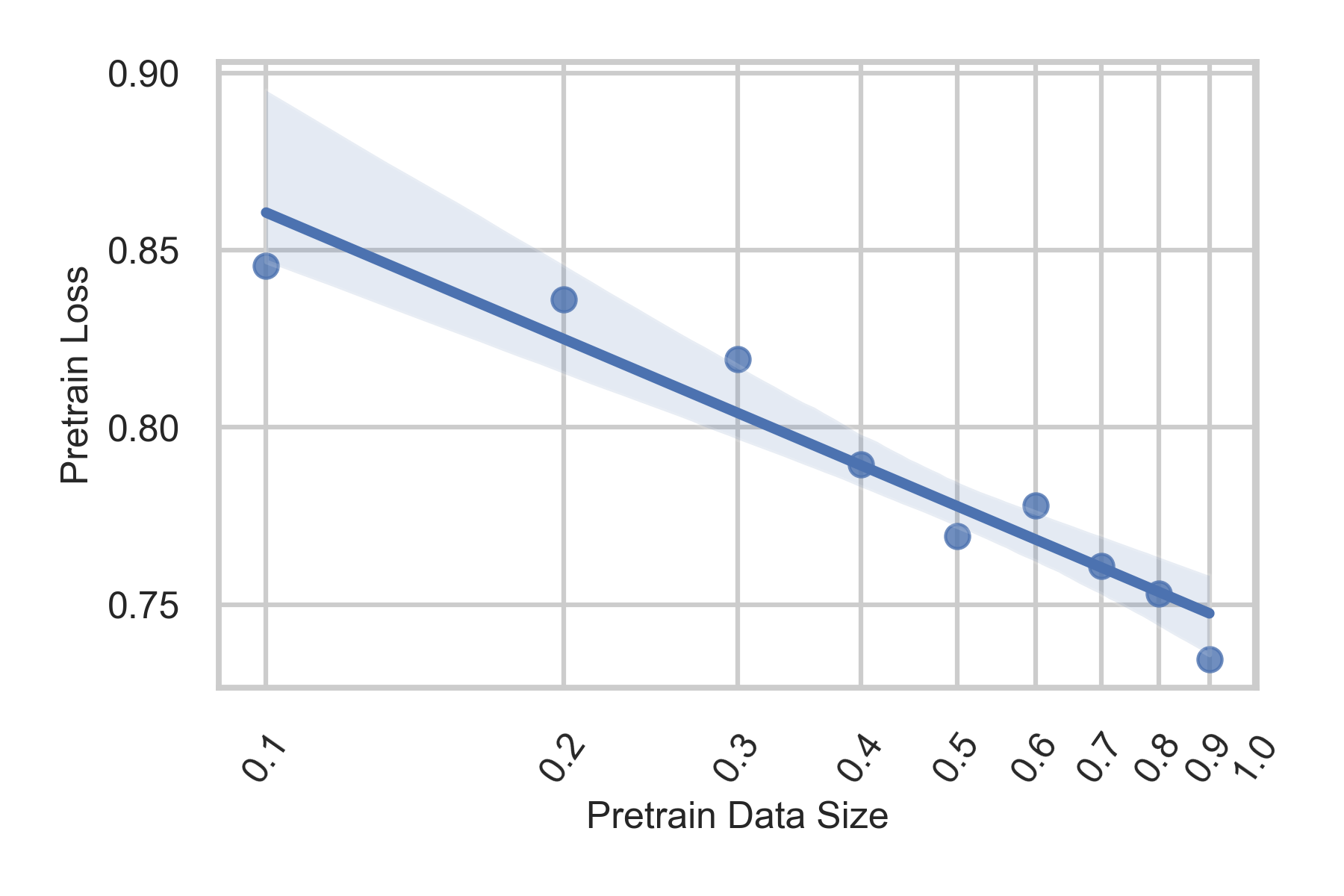}
    \caption{Scaling laws with data ratio $P$ and $\mathcal{L}_p$}
    \label{fig:data-loss}
  \end{subfigure}
  \hfill
  \begin{subfigure}[t]{0.3\textwidth}
    \includegraphics[width=\linewidth]{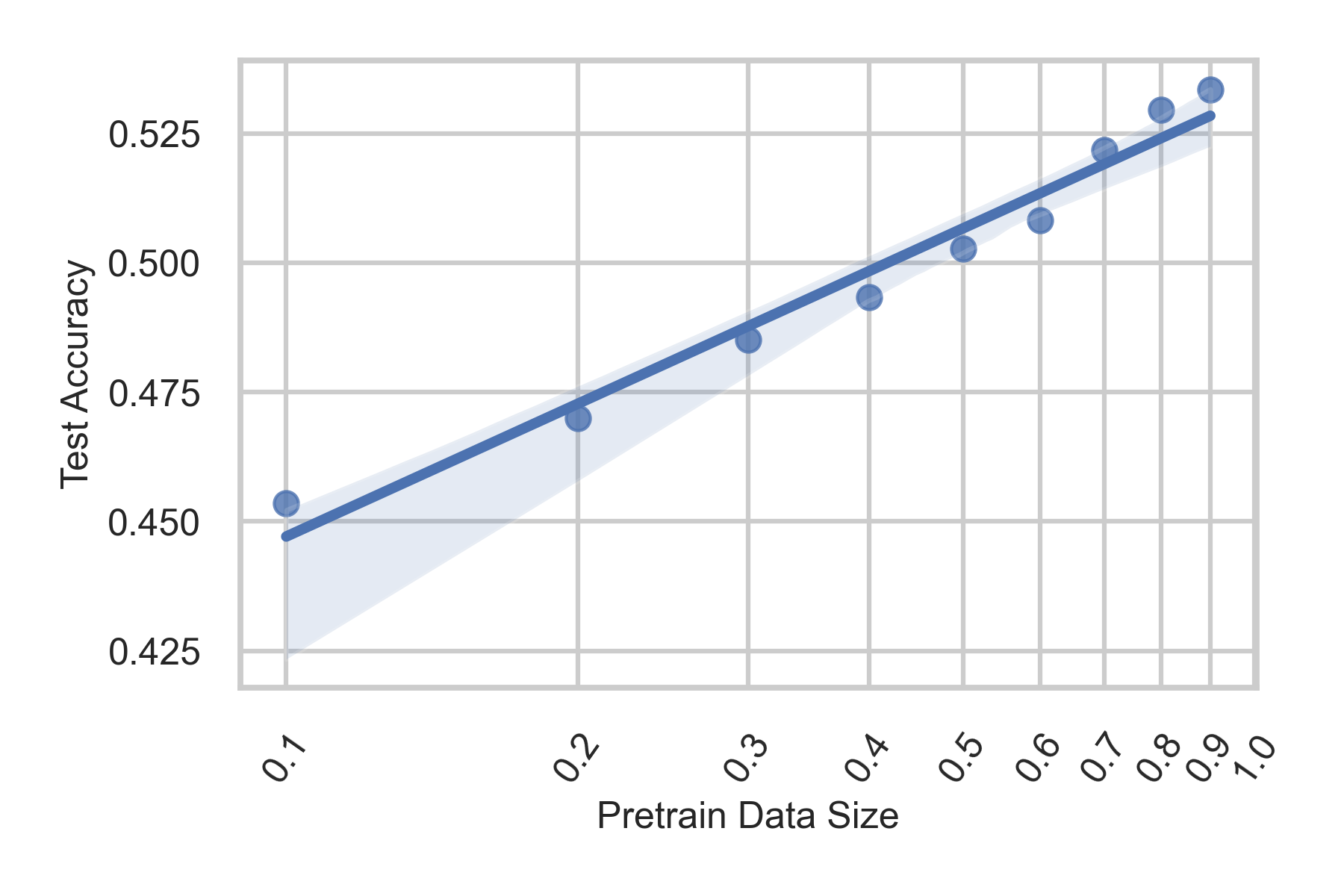}
    \caption{Scaling laws with data ratio $P$ and SEED's balanced accuracy}
    \label{fig:data-seed}
  \end{subfigure}
  \hfill
  \begin{subfigure}[t]{0.3\textwidth}
    \includegraphics[width=\linewidth]{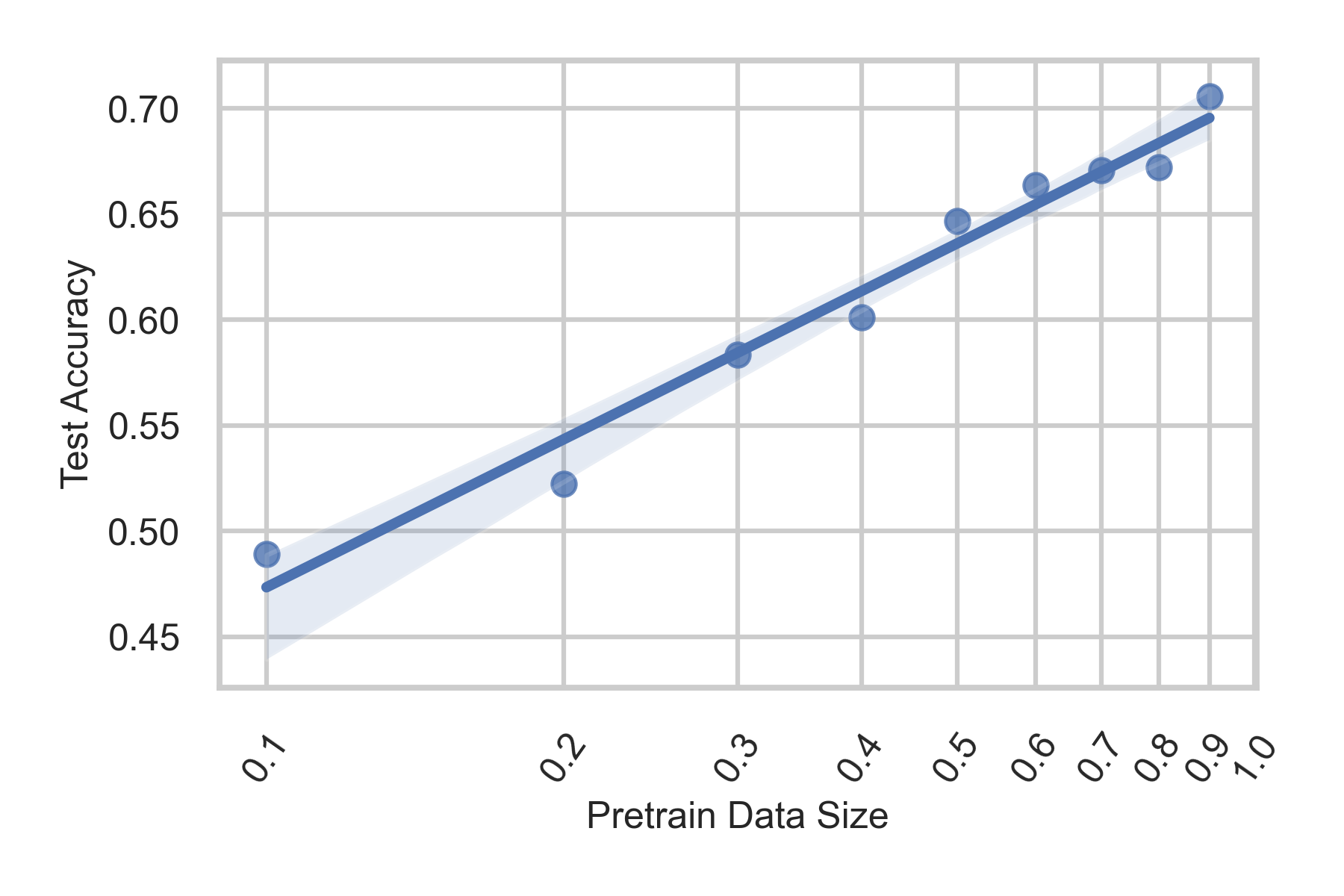}
    \caption{Scaling laws with data ratio $P$ and TUSL's balanced accuracy}
    \label{fig:data-tu}
  \end{subfigure}
  \caption{Scaling laws with data partition $P$ and a) $\mathcal{L}_p$; b) SEED balanced accuracy; c) TUSL balanced accuracy. Axes are all on a logarithmic scale.}
  \label{fig:data}
\end{figure*}

\subsection{For Data Size}

In Figures~\ref{fig:data-loss},~\ref{fig:data-seed},~\ref{fig:data-tu}, we provide the results of the scale law experiments on the pretraining loss function~($\mathcal{L}_p$), SEED, and TUSL dataset.
The results on the pretraining loss function show that the scaling law of the test $\mathcal{L}_p$ with data ratio~($P$) is: \color{black}$\mathcal{L}_p$ = $-0.051$ * $\ln$(P) + $0.742$\color{black}, where $R^2$ is $0.923$.
The results on the SEED dataset show that the scaling law of the test balanced accuracy with data ratio~($P$) is: \color{black}$BAcc$ = $0.037$ * $\ln$(P) + $0.532$\color{black}, where $R^2$ is $0.968$.
The results on the TUSL dataset show that the scaling law of the test balanced accuracy with data ratio~($P$) is: \color{black}$BAcc$ = $0.101$ * $\ln$(P) + $0.706$\color{black}, where $R^2$ is $0.971$.
The results of the pretraining loss and downstream task balanced accuracy indicate that more pretraining data generally achieve higher accuracy.

\section{Limitations and Future Work}\label{sec:limit}

First, although we pretrain the ALFEE with up to 540M parameters using 25,000 hours of EEG data, the overall data scale and model capacity remain limited compared to those of contemporary language and vision models, and the variable quality and low signal-to-noise ratio of EEG data necessitate new preprocessing techniques for more robust representation learning.
Second, while the cross-attention layer combined with task-specific \texttt{CLS} tokens enables excellent multi-task performance under full-parameter fine-tuning, integrating new tasks still requires additional fine-tuning; developing a more comprehensive and universal model similar to those in natural language processing remains an open challenge.
Finally, our current work primarily focuses on robust EEG representation; further research is needed to extend the model to multimodal scenarios such as Video-EEG, Image-EEG, Text-EEG, and fMRI-EEG for broader applicability.

\bibliographystyle{ALFEE}
\bibliography{ALFEE}

\end{document}